\documentclass[sigconf,10pt]{acmart}
\renewcommand\footnotetextcopyrightpermission[1]{} 
\usepackage{graphicx}
\usepackage{times}
\usepackage{balance}
\usepackage{epsfig}
\usepackage{booktabs}
\usepackage{graphicx}
\usepackage{amsmath}
\usepackage{tabu}
\usepackage{subcaption}
\usepackage{amsfonts}
\usepackage{color}
\usepackage{multirow}
\usepackage{algorithm}
\usepackage{algorithmic}
\usepackage[utf8]{inputenc}
\usepackage[T1]{fontenc}
\usepackage{capt-of}
\usepackage{enumerate}
\usepackage{xspace}
\usepackage{pifont}
\usepackage{array}
\usepackage{bm}
\usepackage{csquotes}

\usepackage{url}
\usepackage{hyperref}
\usepackage[hyphenbreaks]{breakurl}
\usepackage[flushleft]{threeparttable}
\usepackage{tablefootnote}

\newcommand{\jc}[1]{{\footnotesize{\color{red}\xspace}}}
\newcommand{\fixed}[1]{{\footnotesize{\color{green} \xspace}}}

\newcommand{\jceditnew}[1]{{\color{black} #1\xspace}}
\newcommand{\editsec}[1]{{\color{black} #1\xspace}}

\newcommand{\camready}[1]{{\color{black} #1\xspace}}

\newcommand{\htedit}[1]{{\color{black} #1\xspace}}

\newenvironment{packed_itemize}{
\begin{list}{\labelitemi}{\leftmargin=0.5em}
  \setlength{\itemsep}{1pt}
  \setlength{\parskip}{0pt}
  \setlength{\parsep}{0pt}
  \setlength{\headsep}{0pt}
  \setlength{\topskip}{0pt}
  \setlength{\topmargin}{0pt}
  \setlength{\topsep}{0pt}
  \setlength{\partopsep}{0pt}  
}{\end{list}}

\newfont{\mycrnotice}{ptmr8t at 7pt}
\newfont{\myconfname}{ptmri8t at 7pt}

\makeatletter
\newcommand{\name}{{\fontsize{9}{9}\selectfont{\textsf{Yoda}}\xspace}\xspace}

\newcommand{\ignore}[1]{\xspace}

\newcounter{packednmbr}
\newenvironment{packedenumerate}{\begin{list}{\thepackednmbr.}{\usecounter{packednmbr}\setlength{\itemsep}{1pt}\addtolength{\labelwidth}{-4pt}\setlength{\leftmargin}{2ex}\setlength{\listparindent}{\parindent}\setlength{\parsep}{1pt}\setlength{\topsep}{2.5pt}}}{\end{list}}

\newenvironment{packeditemize}{\begin{list}{$\bullet$}{\setlength{\itemsep}{1pt}\addtolength{\labelwidth}{-4pt}\setlength{\leftmargin}{2ex}\setlength{\listparindent}{\parindent}\setlength{\parsep}{1pt}\setlength{\topsep}{2.5pt}}}{\end{list}}

\newcounter{insightlabel}
\newcounter{insightnmbr}
\renewcommand{\theinsightlabel}{\textbf{\theinsightnmbr}}

\newcounter{findinglabel}
\newcounter{findingnmbr}
\renewcommand{\thefindinglabel}{\textbf{\thefindingnmbr}}
\newenvironment{finding}{
\vspace{0.2cm}
\begin{list}{\textbf{Finding \thefindinglabel:~}}{\usecounter{findinglabel}\stepcounter{findingnmbr}\setlength{\labelwidth}{0pt}\setlength{\labelsep}{0pt}\setlength{\leftmargin}{0.09in}\setlength{\rightmargin}{0.09in}
\item \em }}{\\[-7pt]\end{list}
}

\newenvironment{statement}{
\vspace{0.2cm}
\begin{list}{\textbf{}}{\setlength{\labelwidth}{0pt}\setlength{\labelsep}{0pt}\setlength{\leftmargin}{0.09in}\setlength{\rightmargin}{0.09in}
\item \em }}{\\[-7pt]\end{list}
}

\newcommand{\tightcaption}[1]{\vspace{-0.1cm}\caption{{\normalfont{\textit{{#1}}}}}\vspace{-0.1cm}}
\newcommand{\tightsection}[1]{\vspace{-0.0cm}\section{#1}\vspace{-0.0cm}}
\newcommand{\tightsubsection}[1]{\vspace{-0.0cm}\subsection{#1}\vspace{-0.0cm}}

\newcommand{\eg}{{\it e.g.,}\xspace}
\newcommand{\ie}{{\it i.e.,}\xspace}

\newcommand{\mypara}[1]{\vspace{0.1cm}\noindent{\bf {#1}:}~}

\newcommand{\myparaq}[1]{\smallskip\noindent{\bf {#1}?}~}

\newcommand{\vp}{VAP\xspace}
\newcommand{\Vp}{VAP\xspace}
\newcommand{\vps}{VAPs\xspace}
\newcommand{\Vps}{VAPs\xspace}

\newcommand{\pc}{PC\xspace}
\newcommand{\Pc}{PC\xspace}

\newcommand{\feature}[1]{{\fontsize{8.5}{8.5}\selectfont\textsf{\em #1}}\xspace}
\newcommand{\featuresmall}[1]{{\fontsize{8}{8}\selectfont\textsf{\em #1}}\xspace}

\newcommand{\Temporal}{\ensuremath{t}\xspace}
\newcommand{\TemporalOracle}{\ensuremath{t^*}\xspace}

\newcommand{\Spatial}{\ensuremath{s}\xspace}
\newcommand{\SpatialOracle}{\ensuremath{s^*}\xspace}

\newcommand{\Class}{\ensuremath{m}\xspace}
\newcommand{\ClassOracle}{\ensuremath{m^*}\xspace}

\newcommand{\Pipeline}{\ensuremath{v}\xspace}
\newcommand{\Feature}{\ensuremath{f}\xspace}
\newcommand{\FeatureVec}{\mathbb{F}\xspace}

\newcommand{\FeatureValVec}{\ensuremath{x}\xspace}
\newcommand{\Perf}{\mathbb{P}\xspace}

\AtBeginDocument{%
  \providecommand\BibTeX{{%
    \normalfont B\kern-0.5em{\scshape i\kern-0.25em b}\kern-0.8em\TeX}}}

\acmYear{2021}\copyrightyear{2021}
\setcopyright{acmcopyright}
\acmConference[SEC '21]{SEC '21: 2021 IEEE/ACM Symposium on Edge Computing (SEC)}{December 14--17, 2021}{San Jose, CA, USA}
\acmBooktitle{SEC '21: 2021 IEEE/ACM Symposium on Edge Computing (SEC), December 14--17, 2021, San Jose, CA, USA}
\acmPrice{15.00}
\acmDOI{10.1145/3341302.XXXXXXX}
\acmISBN{XXX-X-XXXX-XXXX-X/XX/XX}

\makeatother

\begin{document}
\pagestyle{plain} 

\title{\scalebox{0.975}{Towards Performance Clarity of Edge Video Analytics}}

\author{Zhujun Xiao}
\affiliation{\institution{University of Chicago}
  \city{Chicago}
  \state{Illinois}
  \country{USA}}
\email{zhujunxiao@cs.uchicago.edu}

\author{Zhengxu Xia}
\affiliation{\institution{University of Chicago}
  \city{Chicago}
  \state{Illinois}
  \country{USA}}
\email{zxxia@uchicago.edu}

\author{Haitao Zheng}
\affiliation{\institution{University of Chicago}
  \city{Chicago}
  \state{Illinois}
  \country{USA}}
\email{htzheng@cs.uchicago.edu}

\author{Ben Y. Zhao}
\affiliation{\institution{University of Chicago}
  \city{Chicago}
  \state{Illinois}
  \country{USA}}
\email{ravenben@cs.uchicago.edu}

\author{Junchen Jiang}
\affiliation{\institution{University of Chicago}
  \city{Chicago}
  \state{Illinois}
  \country{USA}}
\email{junchenj@cs.uchicago.edu}
\renewcommand{\shortauthors}{Zhujun Xiao, et al.}


\begin{abstract}
Edge video analytics is becoming the solution to many safety
and management tasks. Its wide deployment,
 however, must first address the tension between inference accuracy and
 resource (compute/network) cost.   This \jceditnew{has} led to the development of
 video analytics pipelines (VAPs), which reduce resource cost
 by combining DNN compression/speedup techniques with video
 processing heuristics. Our measurement study, however, shows that today's
methods for evaluating VAPs are incomplete, often producing premature
conclusions or ambiguous results.  This is because each VAP's
performance varies substantially across videos and time, and 
\jceditnew{is sensitive to}
different subsets of video content characteristics.

We argue that accurate VAP evaluation must
first characterize the complex interaction between VAPs and video 
characteristics,  which we refer to as VAP performance clarity. 
We design and implement \name, the first VAP benchmark to achieve performance
clarity.  Using primitive-based profiling and a carefully curated
benchmark video set, \name
builds a performance clarity profile for each VAP to precisely define its
accuracy/cost tradeoff and its relationship with video
characteristics.  We show that \name substantially improves VAP evaluations
by (1) providing a comprehensive, transparent
assessment of VAP performance and its dependencies on video
characteristics; (2) \jceditnew{explicitly identifying} fine-grained VAP 
  behaviors that were previously hidden by large performance
  \jceditnew{variance}; and (3) revealing strengths/weaknesses among different
  VAPs and new design opportunities.

\end{abstract}



\maketitle


\tightsection{Introduction}

Edge video analytics is becoming the modern solution
to many critical tasks~\cite{videoanalyticsapp}.  With the ability
to accurately detect, recognize and track objects on the fly,   it can quickly detect
and respond to traffic accidents and hazard events~\cite{trafficvision, traffictechnologytoday, goodvision,
  intuvisiontech, vision-zero,
    visionzero-bellevue, msr-bellevue},   monitor and enforce physical distance during
COVID-19~\cite{cameracovid,cameracovid1}, auto-manage retail
stores and factories~\cite{smartretail}, and perform surveillance functions to make the
world safer~\cite{ChicagoCam,LondonCCTV}.

Deployment of edge video analytics at scale, however, must address the
tension between inference accuracy
and resource cost, i.e., {\em compute} cost to run
inference tasks and/or {\em bandwidth} cost to transfer data from cameras to
servers~\cite{chen2019deep,noghabi2020emerging}.  This tension continues to grow as video sources proliferate at the
network's
edge~\cite{ChicagoCam,LondonCCTV,VideoAnalyticsMarket,slate-video-news,infowatch16},
separated from the heavy compute power necessary to run large deep neural networks (DNNs)
by a bandwidth-constrained mobile network.

\begin{figure}[h]
\centering
\includegraphics[width=0.4\textwidth]{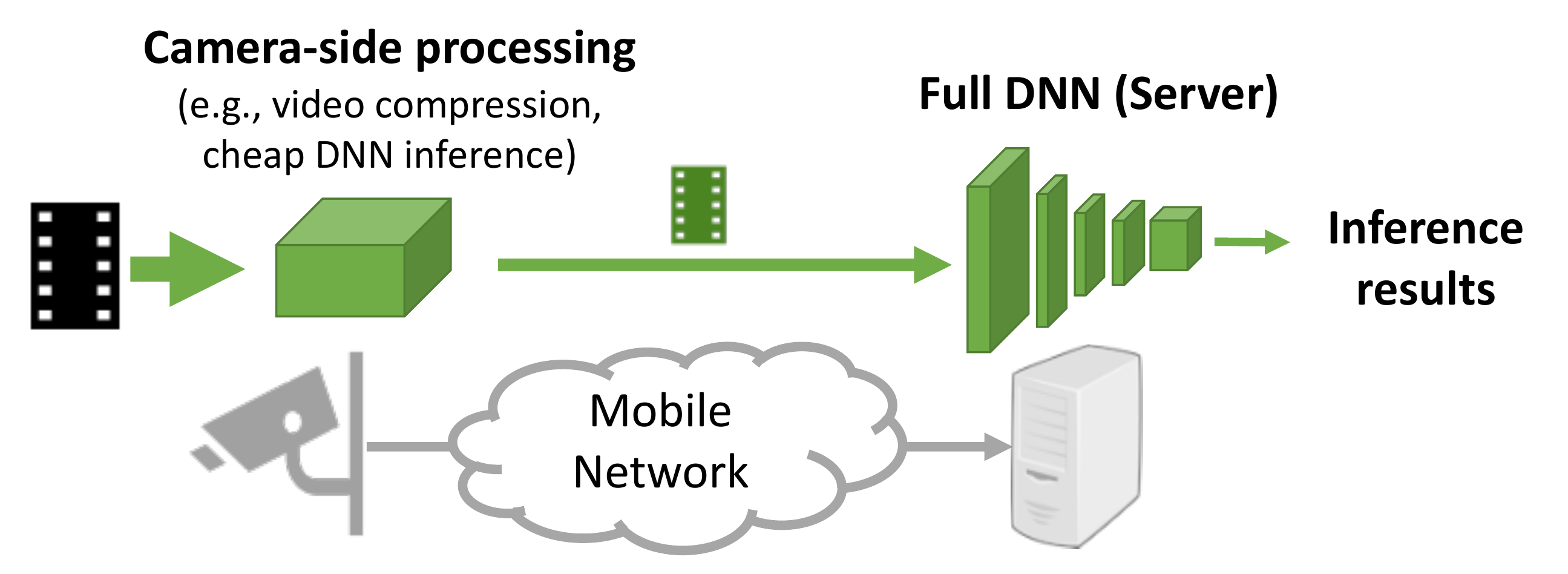}
\tightcaption{Illustration of a video analytics pipeline (\vp).}
\vspace{-0.1cm}
\label{fig:pipeline-intro}
\end{figure}

In response, researchers have developed numerous 
{\em video analytics pipelines {(\vps)}} to optimize the accuracy and cost
tradeoff~\cite{noscope,chameleon,fast,videostorm,mainstream,videoedge,mcdnn,glimpse,awstream,dds-hotcloud,rexcam-hotmobile,vigil,emmons2019cracking,eaar},
by combining DNN model compression/speedup techniques with video processing heuristics
such as frame sampling and image downsizing (see Figure~\ref{fig:pipeline-intro}). For instance, 
Chameleon~\cite{chameleon} shows that intelligently subsampling
traffic video frames at the cameras can effectively reduce network and compute costs
without degrading inference accuracy.

As edge video analytics
and \vps continue to evolve, accurate and transparent evaluation of
\vps 
becomes critical. 
\editsec{For instance, operators of edge video analytics need to know 
what the optimal \vp is for a given video input, how often the
network/compute usage exceeds a budget, or how often accuracy drops
below a threshold.}


\camready{
 \mypara{Evaluating VAPs} Today, \vps are evaluated using some corpus
 of past video samples that represent the target scenario(s).  After running \vps on
 these videos, their performance (i.e., the accuracy and cost
 tradeoff) is analyzed and compared against each other.  
 Following this method, we run an empirical study to evaluate seven
\vps from recent papers, using a large chunk (14.5 hours) of traffic videos. 
Our study shows that today's evaluation method is insufficient to characterize \vps, often
leading to partial/premature conclusions on the efficacy of a \vp and across \vps. This is because \vp performance has a
strong dependency on video content -- it can vary substantially across
videos even in the same scenario
({\em e.g.\/}, highway traffic cameras), and drift dramatically over
time when operating on the same camera. Therefore,  today's evaluation is either biased by the use of short video clips or
produces vague results over long videos, i.e., an
excessively wide distribution of possible cost-accuracy outcomes. 
}




Our measurement study suggests that an ideal evaluation of \vps must 
\camready{have high performance coverage and low performance variance. 
Here,  ``high coverage'' means the evaluation reveals both good and bad performance of a \vp,
whereas ``low variance'' means the evaluation could accurately estimate the \vp's performance on 
individual videos.
And the strong dependency of \vp performance on video content suggests such ideal evaluation must} 
\editsec{characterize
the complex interactions between video workloads and a \vp's performance}.  Doing so presents
three distinct benefits for \vp design and deployment: (1) providing a
comprehensive assessment of \vps under diverse video characteristics; (2)
understanding how/why each \vp's performance varies across videos; (3)
revealing relative strengths among \vps under different video content
characteristics.  We refer to this new evaluation requirement
as {\em \vp\ performance clarity}.

\mypara{Achieving performance clarity} A direct approach would test \vps
exhaustively on a large collection of mobile video 
workloads, {\em e.g.\/} {\em existing} video collections developed for testing DNN
models~\cite{imagenet,coco,google-benchmark,
  tbd,dawnbench,visual-road,coleman2019analysis}.  Yet these are designed to
evaluate DNN architectures rather than \vps, thus lack sufficient coverage of
video characteristics that will affect \vp performance.  An alternative is to build a database of empirical
workloads that covers all possible video feature value combinations, and use
them to test \vps. This is intractable, however, since it would require a large database
capturing an exponential number of video feature combinations.

Instead, we propose to characterize {\vp} performance using a carefully
curated set of videos that serve to evaluate different aspects of \vps.  Our design is based
on the observation
that each \vp is inherently modular and can be
broken into a set of  ``global'' 
primitives. Each primitive leverages a distinct set of video processing 
heuristics to optimize the accuracy/cost
tradeoff, and thus can be profiled independently (against its associated
video features) and then (re)assembled to profile full 
\vps.   This modular structure allows us to efficiently 
profile each full \vp by
combining its corresponding primitive-specific profiles. 
Note that some prior works also observe independent
\vp modules but use it to refine particular \vp designs~\cite{videoedge,chameleon}.
In contrast, we leverage this observation to design 
accurate evaluation of many \vps.

We present {\bf \name, the
first \vp benchmark} designed to achieve performance clarity.  Using
a carefully curated set of benchmark videos (67 minutes in length), \name focuses on characterizing the
complex dependencies of \vp performance on mobile video content
characteristics, and does so efficiently.  For each \vp $v$, \name builds
a performance clarity profile $(\mathbb{P}_v)$ by running $v$ on a set of benchmark videos parameterized by a
set of video features, both chosen based
on $v$'s design primitives.  The resulting $\mathbb{P}_v$ is a lookup
table that lists $v$'s performance (the accuracy/cost relationship)
under different video feature values.   This provides a comprehensive and
transparent  assessment of $v$'s performance and its dependencies on
video features. 
We show \name's contributions towards \vp
evaluation,  in three concrete aspects.

\begin{figure*}[t]
\centering
\includegraphics[width=0.98\textwidth]{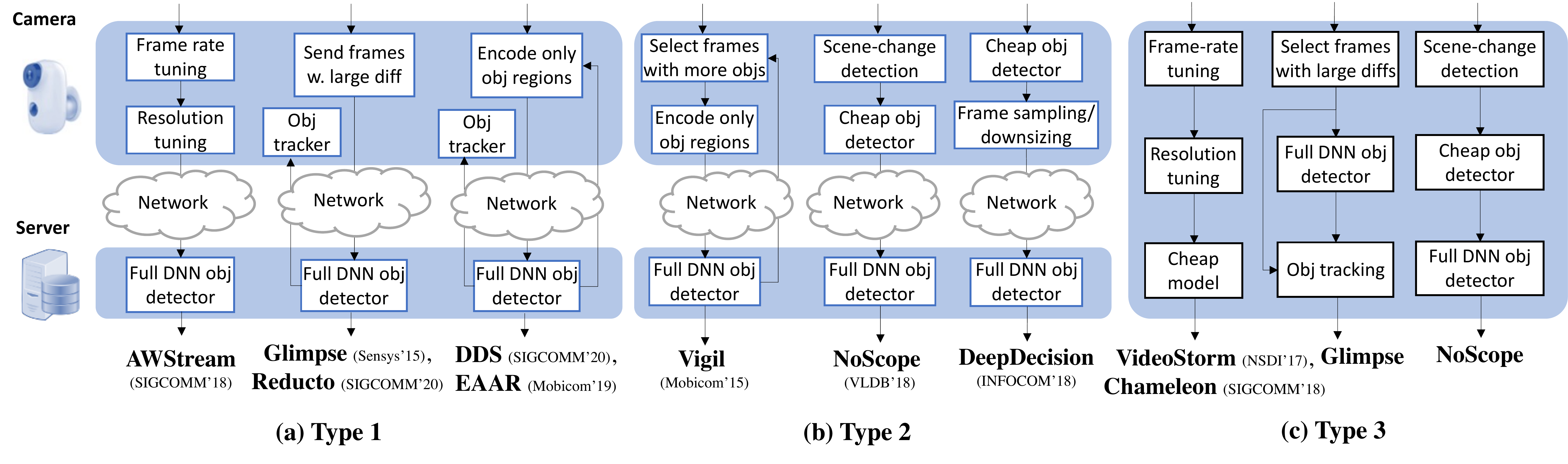}
\vspace{-0.2cm}
\tightcaption{Schematic illustration of some example {\vps} grouped into
  three general types. (Differences within each general scheme are omitted here.) \editsec{Our goal is not to list all \vps; instead, we seek to identify common techniques and their performance.}}
\label{fig:pipelines-illustrated}
\end{figure*}


\begin{packed_itemize} \vspace{-0.03in}
\item {\bf Performance clarity} -- \name accurately captures
  existing \vps' performance and their dependencies on video
  features.
  It  largely outperforms existing \vp evaluations with higher coverage 
    (the completeness of the evaluation) and lower variance (the
    ambiguity of the evaluation outcome).

\item  {\bf Performance prediction} -- 
Using $\mathbb{P}_v$, \name can efficiently estimate $v$'s performance for videos {\em not} included in the benchmark set, without running $v$.
This takes 2 orders of magnitude less computation than 
running $v$ on the video.

\item {\bf Practical insight for \vp deployment} -- 
\name's \vp profiles expose strengths and
  weaknesses among existing \vps, and the underlying deployment 
  scenarios and video features associated with these
  conclusions. These insights allow us to identify previously hidden gaps and
  opportunities to guide/motivate future VAP designs. \vspace{-0.03in}
 \end{packed_itemize}

Though \name serves well on the seven \vps considered in this work, it is {\em not} without limitations.
Currently, \name's content features and benchmark videos are
not future-proof (\eg \name does not support multi-stream/multi-query
\vps).  For distributed \vps that handle bandwidth-constrained
connections, \name only evaluates reductions in average network
bandwidth usage
but not the impact of bandwidth fluctuation.

Nonetheless, as the first attempt at benchmarking \vps' performance clarity, \name suggests a viable path towards profiling the dependencies of \vps' performance on video content via a {\em modularized} approach.
Our goal is not to realize an ``ideal'' benchmark;
rather, we provide a concrete implementation of the proposed benchmark, which validates the need for performance clarity and initial
 feasibility on accurate performance evaluations of \vps,  and provides new 
 insights for \vp design and deployment.  We release the \name
 toolkit \camready{in https://yoda.cs.uchicago.edu} and plan to
 expand our study to include other \vps and additional video features.


\tightsection{Background}
\label{sec:background}
In this section, we present an overview on existing \vps,
focusing on their design objectives and evaluations.

\subsection{VAP Design} 
Computer-vision DNNs are generally optimized for high accuracy.
However, the compute and network cost to achieve such accuracy can be high\footnote{For instance, running state-of-the-art object detector at 30fps requires one NVidia GTX Titan X GPU (>\$1.1K)~\cite{google-benchmark} and streaming the video at 720p ($\sim5$Mbps) costs \$2K/day for AT\&T 4G LTE network (\$50 for 30GB data before the speed drops to a measly 128kbps~\cite{4G-plan}).}. 
This tension between accuracy and cost
has stimulated many ongoing efforts to develop video analytics pipelines~(\vps)~\cite{awstream,chameleon,vigil,eaar,videostorm,mainstream,noscope,reducto,dds-sigcomm}.  \vps reduce network/compute cost while maintaining
high inference accuracy,  by combining DNN
compression/speedup methods and video processing heuristics such as frame sampling and image
downsizing.

Existing \vps fall in three general types (Figure~\ref{fig:pipelines-illustrated}).

\begin{packed_itemize}
\item {\bf Type 1: Saving network cost when the camera has low local compute power.}
The camera only encodes video frames and runs simple tracking algorithms, but does not perform any inference that requires accelerators such as GPUs.
Instead, a \vp saves network cost by selecting a subset of frames/pixels to send to the server for DNN inference.
For example,  {\em AWStream}~\cite{awstream} adapts video frame rate,
resolution and quality.
{\em Glimpse}~\cite{glimpse} and {\em Reducto}~\cite{reducto} send only frames that contain new objects (\eg identified by measuring inter-frame  difference).
Similarly, {\em EAAR}~\cite{eaar} and {\em DDS}~\cite{dds-sigcomm} only encode regions that are likely relevant to the inference task.

\item {\bf Type 2: Saving network cost when the camera and the server
    split the inference task.}  Here the camera device is equipped with some inference power (\eg with
  a low-power GPU) and thus can run a cheap DNN.  For example, {\em Vigil}~\cite{vigil} runs a cheap
  object detector on the camera to identify regions containing
  most objects and sends only these regions to the server for full DNN
  inference. {\em NoScope}~\cite{noscope} first identifies frames with significant
pixel changes and runs a cheap DNN (fine-tuned per video
stream) on these frames. Only when the cheap DNN has low
confidence will the frames be sent to the server for further inference.

\item{\bf Type 3: Saving compute cost of a resource-constrained edge device.}
The third type of \vps reduces compute cost, when a camera device (or edge server)
has moderate compute power to run some inference
locally.
{\em Videostorm}~\cite{videostorm} and {\em Chameleon}~\cite{chameleon} uniformly sample frames, downsize the sampled frames to a lower resolution, and process them using a less accurate yet cheaper DNN model.
We note that {\em Glimpse} (Type 1),
{\em NoScope} (Type 2) can also be applied here
to reduce compute cost, and thus fall into this type.
\end{packed_itemize}

\camready{
\subsection{How Are \vps Evaluated Today?} 
\label{subsec:prior-eval}
Today's evaluation empirically tests and compares the \vps'
performance (accuracy, cost) on a set of videos collected from the
target scenario(s)~\cite{noscope,videostorm,videoedge,chameleon}, \eg some traffic videos recorded by fixed cameras in
urban crossroads. 
Table~\ref{tab:eval} lists the target scenarios and videos
(sources and lengths) used to evaluate some recent
\vps.
}

Such evaluation relies on an implicit assumption:
\begin{statement}
{\bf Today's evaluation assumption:} A \vp's performance under a target scenario can be
  represented by its performance seen on a set of
  long videos of the same scenario.
\end{statement}
\noindent Unfortunately, this is not always true.  Our own measurement
study shows that a \vp's performance can vary
dramatically among videos of the same scenario (see
\S\ref{subsec:variability}).

\begin{table}[t!]
\footnotesize
\begin{tabular}{p{1.55cm}p{4.2cm}p{1.77cm}}
\hline
 \textbf{\vp} & \textbf{Target scenarios (sources of videos)} ``YT'' =
                YouTube, ``P'' = proprietary & \textbf{Total duration ({\em \# of videos})} \\ \hline\hline
Glimpse~\cite{glimpse} & Moving traffic cams (YT) + Face (P) & 65min ({\em 30})\\\hline
AWStream~\cite{awstream} & Fixed traffic cams (MOT16) + AR (P) & 6.3min ({\em 4}) \\\hline
Vigil~\cite{vigil} & Campus cams + Indoor (P) & 3min ({\em 3}) \\\hline
Reducto~\cite{reducto} & Fixed traffic cams (YT) & 250min (25) \\\hline
Chameleon~\cite{chameleon} & Fixed traffic cams + Indoor (P) & 525min ({\em 15}) \\\hline
DDS~\cite{dds-sigcomm} & Fixed \& Moving traffic cams (YT) & 30.7min ({\em 16}) \\\hline
\end{tabular}
\vspace{0.1cm}
\caption[]{\normalfont{\textit{\editsec{Today, \Vps are evaluated on
        videos of one or two scenarios as a whole. For consistency, we only list object detection datasets.  
}}}\vspace{-0.3cm}}
\label{tab:eval}
\end{table}





\camready{
  \tightsection{Our Empirical Study on \vp Evaluation}
  \label{sec:motivation}

As video analytics and \vps continue to evolve, accurate and
  transparent evaluation of \vps is crucial to their real-world
  adoption. In this work, we are interested in
  understanding whether today's \vp evaluation methods (\S\ref{subsec:prior-eval}) can fulfill
  this requirement.  Since existing \vp proposals generally 
  run evaluation using different datasets, one cannot directly assess 
  and compare their performance from their reported results. 
  Instead, our empirical study evaluates 7 popular \vp designs using the same video datasets (14.5 hours in total) that consist of a much larger and more diverse collection of traffic videos.
 Our analysis reveals significant \vp performance
 variability across videos of the same target scenario, suggesting that today's evaluation method is insufficient to characterize \vps. 
We then discuss its implications for a better \vp evaluation, which
lead to the development of \name.


\subsection{Methodology and Dataset}
\label{subsec:dataset}
We start by discussing the methodology behind our measurement study. 

\mypara{VAPs studied} 
We study and compare the performance of 7 recent \vps on the task of {\em
  object detection}. These include AWStream~\cite{awstream}, Glimpse~\cite{glimpse}, Vigil~\cite{vigil},
NoScope~\cite{noscope}\footnote{We include NoScope in our study since it is also
  applicable to object detection, although it was only evaluated on
  binary classification.}, Videostorm~\cite{videostorm}, Reducto~\cite{reducto}, and DDS~\cite{dds-sigcomm}.  
They cover a wide range of today's \vp design techniques illustrated in 
Figure~\ref{fig:pipelines-illustrated}.


For consistency, we configure all these \vps to operate on videos of (30fps, 720p) and
all use the same pre-trained DNN model as their {\em full DNN model}.
To choose the full DNN model, we experimente with several popular
choices (e.g., FasterRCNN-ResNet101~\cite{google-model-zoo}, 
Yolo~\cite{yolov3}) and select FasterRCNN-ResNet101 since it produces the
highest accuracy in object detection.  Later we also repeat our
experiments using Yolo, and find that while the absolute VAP
performance varies slightly, the key findings remain the same.  
Finally, we consider the scenario where \vps are ``optimally
configured'' to eliminate potential inconsistency or errors introduced
by imperfect system configuration.  For each video segment ($\approx$30s), we configure each \vp by picking its best
parameter values (\eg frame sampling rate of VideoStorm, or inter-frame difference threshold of Glimpse) 
that minimize cost while
achieving over 0.9 inference accuracy in the first 1/3 of the
segment. We  then test and report the \vp performance on the rest of
the video segment.  We believe this consideration helps increase the
fairness and transparency of our \vp evaluation.

\mypara{Our ``coverage'' dataset} 
To show a more complete picture of \vp performance, we compile a {\em
  coverage set} of public traffic videos from a diverse video sources at a much larger
scale than existing works. We target specifically traffic videos since
they are commonly used in \vp evaluation (see Table~\ref{tab:eval}). When
compiling our dataset, we seek to include public traffic videos from diverse
sources, covering different scenarios (fixed or moving cameras;
day or night; highway, city or rural streets), and videos 
displaying a wide range of content characteristics and dynamics, e.g., object
speeds, sizes, object arrival rate.

With these in mind, our final
coverage set consists of 14.5 hours of traffic videos from
multiple sources:  YouTube (32 long videos, 10-47 minutes each), 
Waymo~\cite{sun2019scalability} (5 hours), KITTI~\cite{Geiger2012CVPR}
(20 minutes), and MOT~\cite{MOT16} (8 minutes).  
All videos are split into 2112 segments ($\approx$30s per segment).
}

\begin{figure*}[t]
\centering
\includegraphics[width=1.01\textwidth]{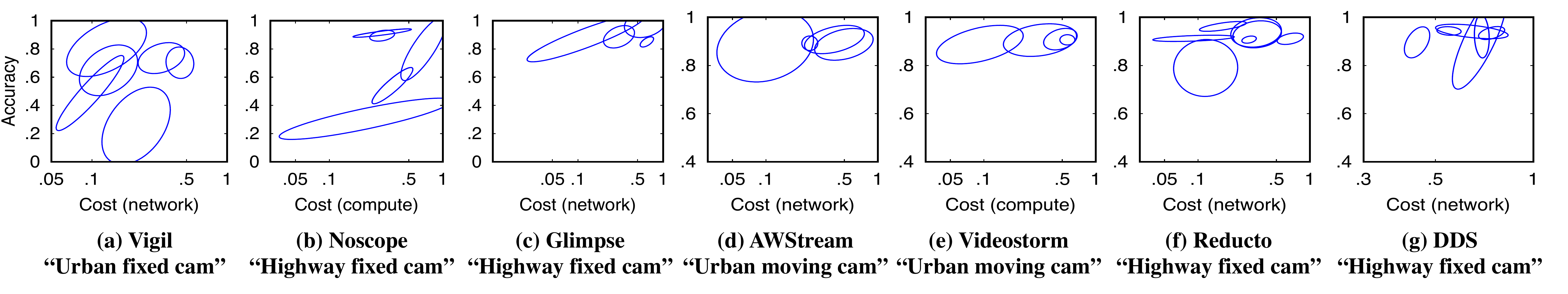}
\vspace{-0.8cm}
\tightcaption{Significant performance variability of the same \vp among videos of the same scenario.
       Each ellipse
   outlines the 1-$\sigma$ range of \vp performance across
       the segments of a video.\vspace{-0.03cm}}
\label{fig:existing-coverage}
\end{figure*}

\begin{figure*}[t!]
\centering
\hspace{-0.42cm}
\begin{subfigure}[b]{0.175\textwidth}
    \includegraphics[width=\linewidth]{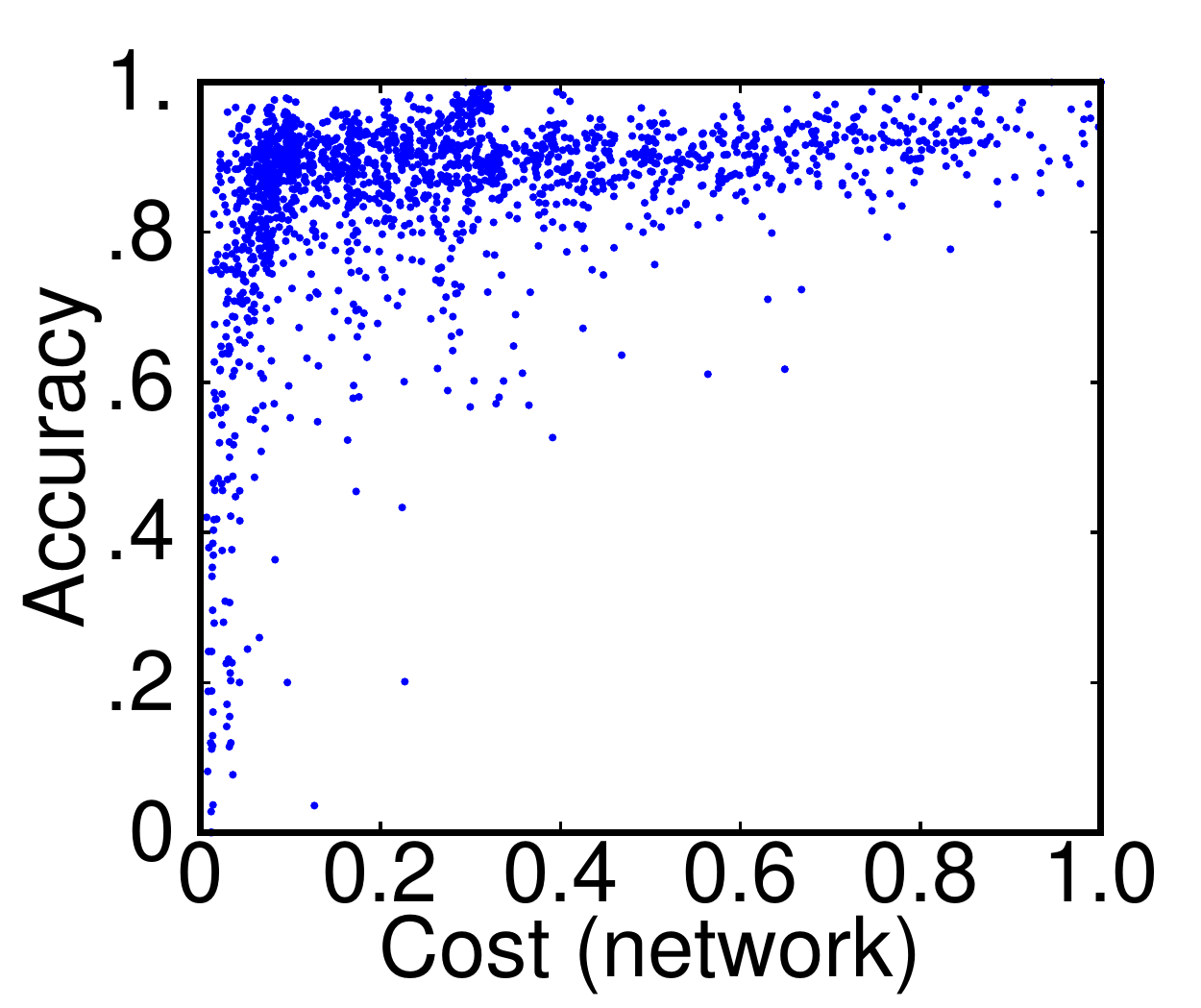}
    \vspace{-0.6cm}
    \subcaption{AWStream (Type 1)}
\end{subfigure}
\hspace{-0.4cm}
\begin{subfigure}[b]{0.175\textwidth}
    \includegraphics[width=\linewidth]{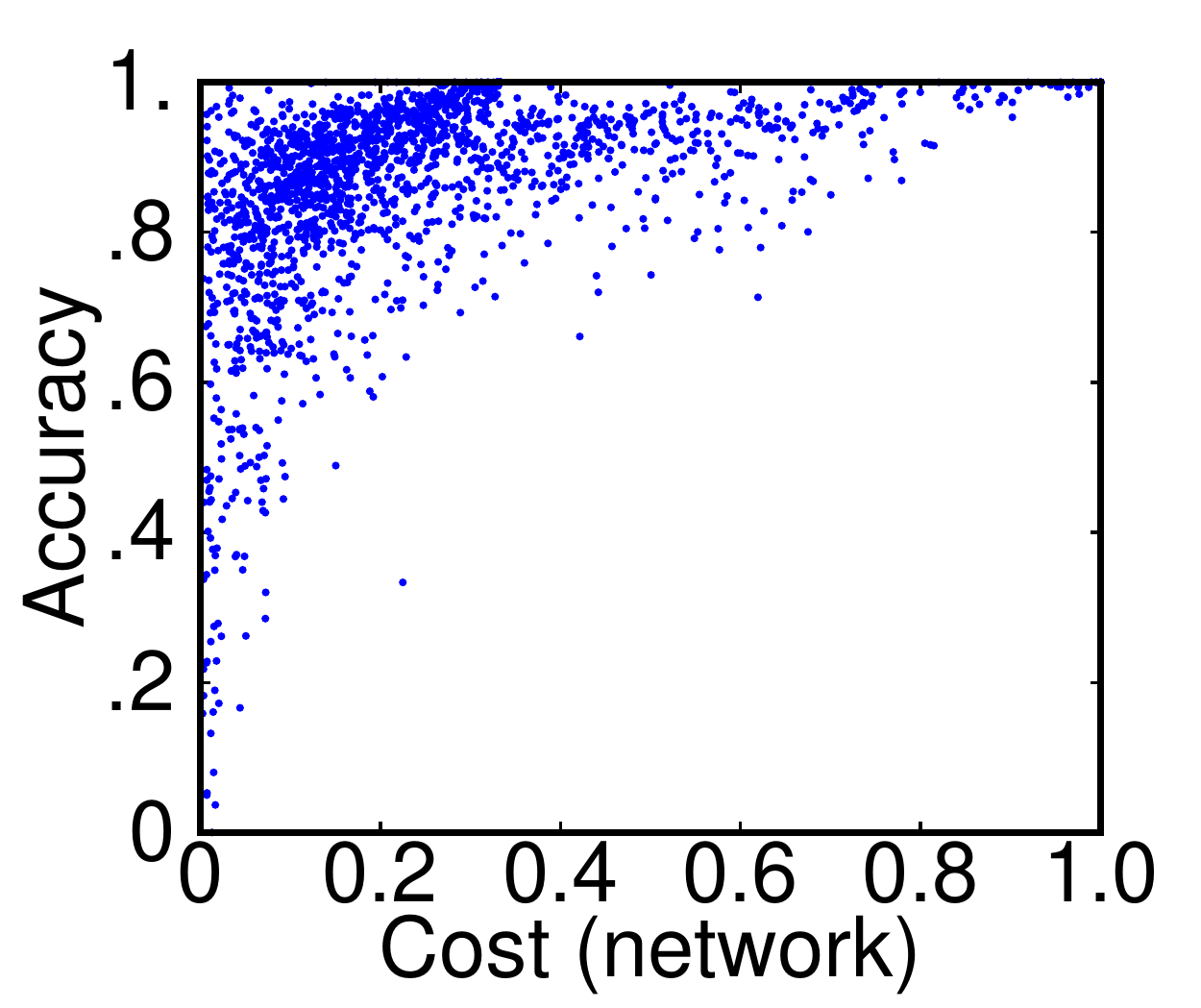}
    \vspace{-0.6cm}
    \subcaption{Glimpse (Type 1)}
\end{subfigure}
\hspace{-0.4cm}
\begin{subfigure}[b]{0.175\textwidth}
    \includegraphics[width=\linewidth]{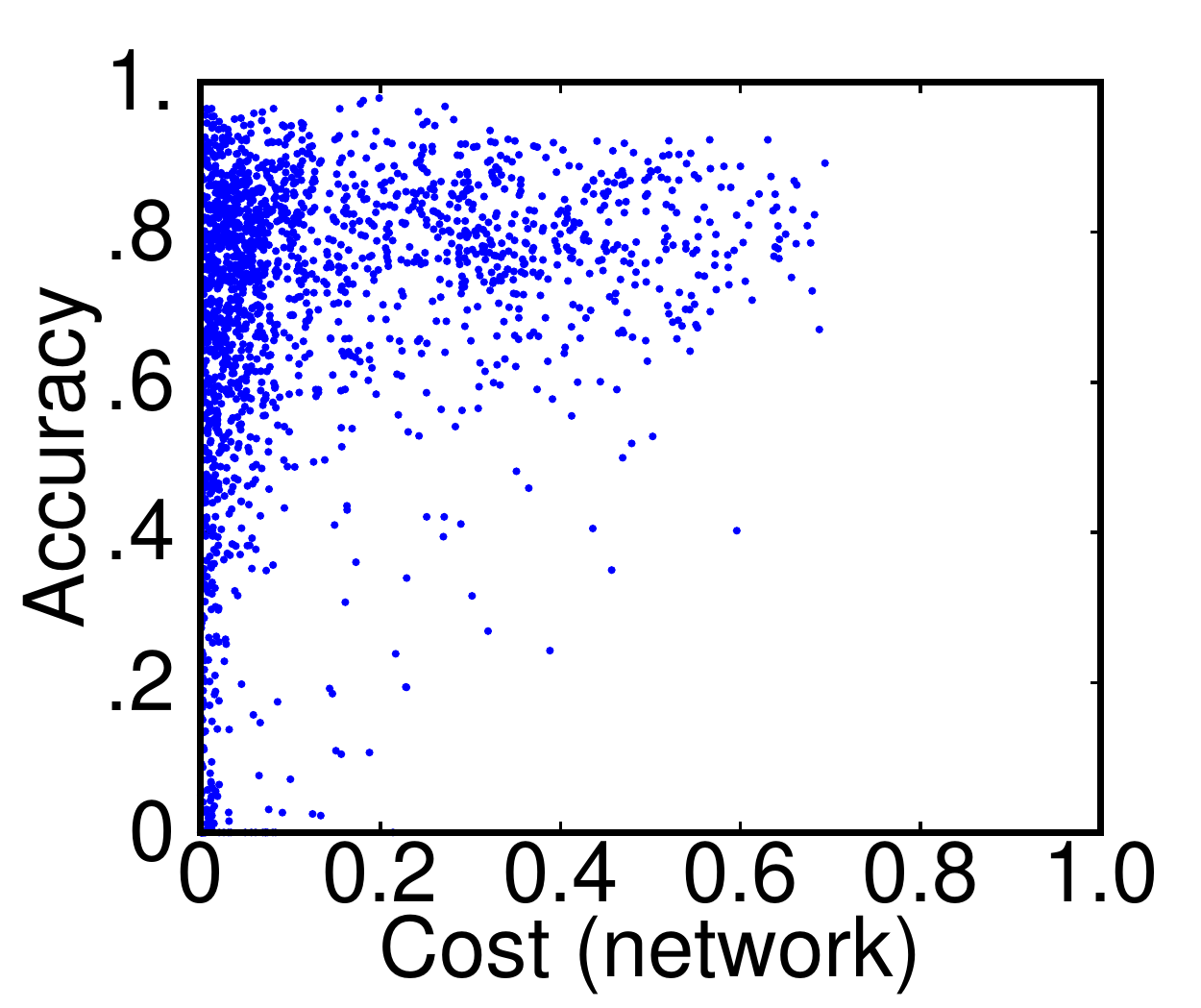}
    \vspace{-0.6cm}
    \subcaption{Vigil (Type 2)}
\end{subfigure}
\hspace{-0.4cm}
\begin{subfigure}[b]{0.175\textwidth}
    \includegraphics[width=\linewidth]{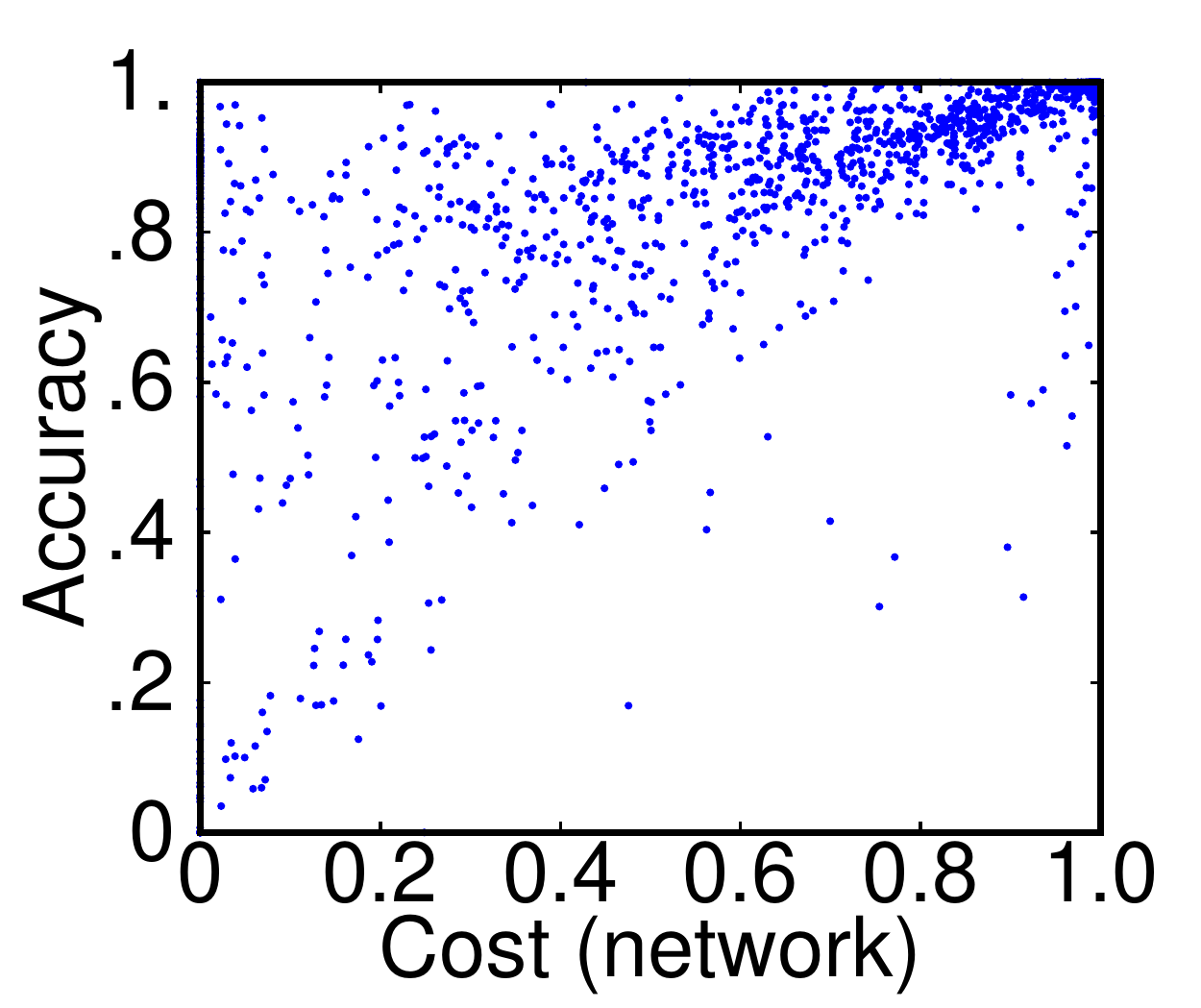}
    \vspace{-0.6cm}
    \subcaption{NoScope (Type 2)}
\end{subfigure}
\hspace{-0.4cm}
\begin{subfigure}[b]{0.175\textwidth}
    \includegraphics[width=\linewidth]{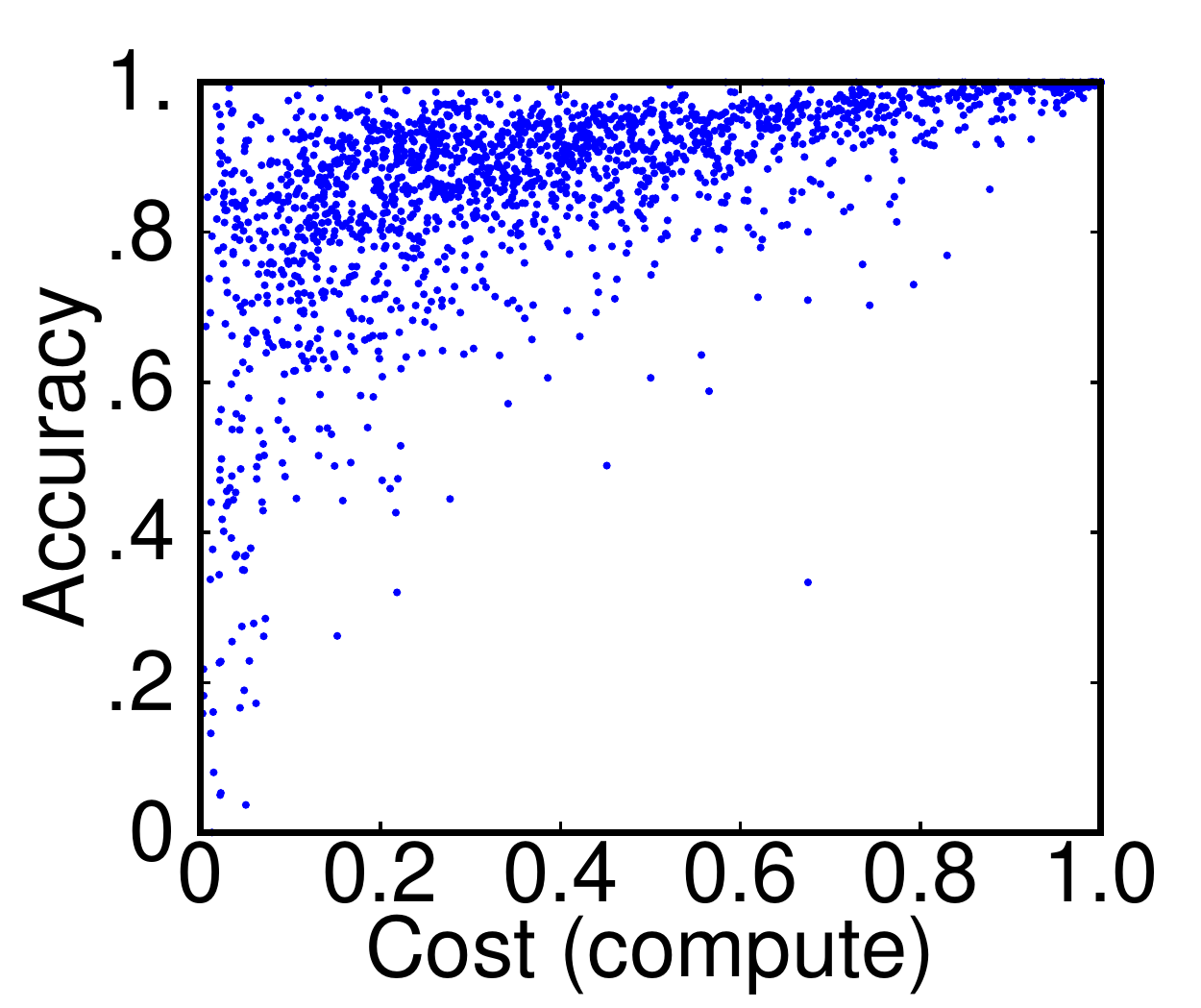}
    \vspace{-0.6cm}
    \subcaption{Glimpse (Type 3)}
\end{subfigure}
\hspace{-0.4cm}
\begin{subfigure}[b]{0.175\textwidth}
    \includegraphics[width=\linewidth]{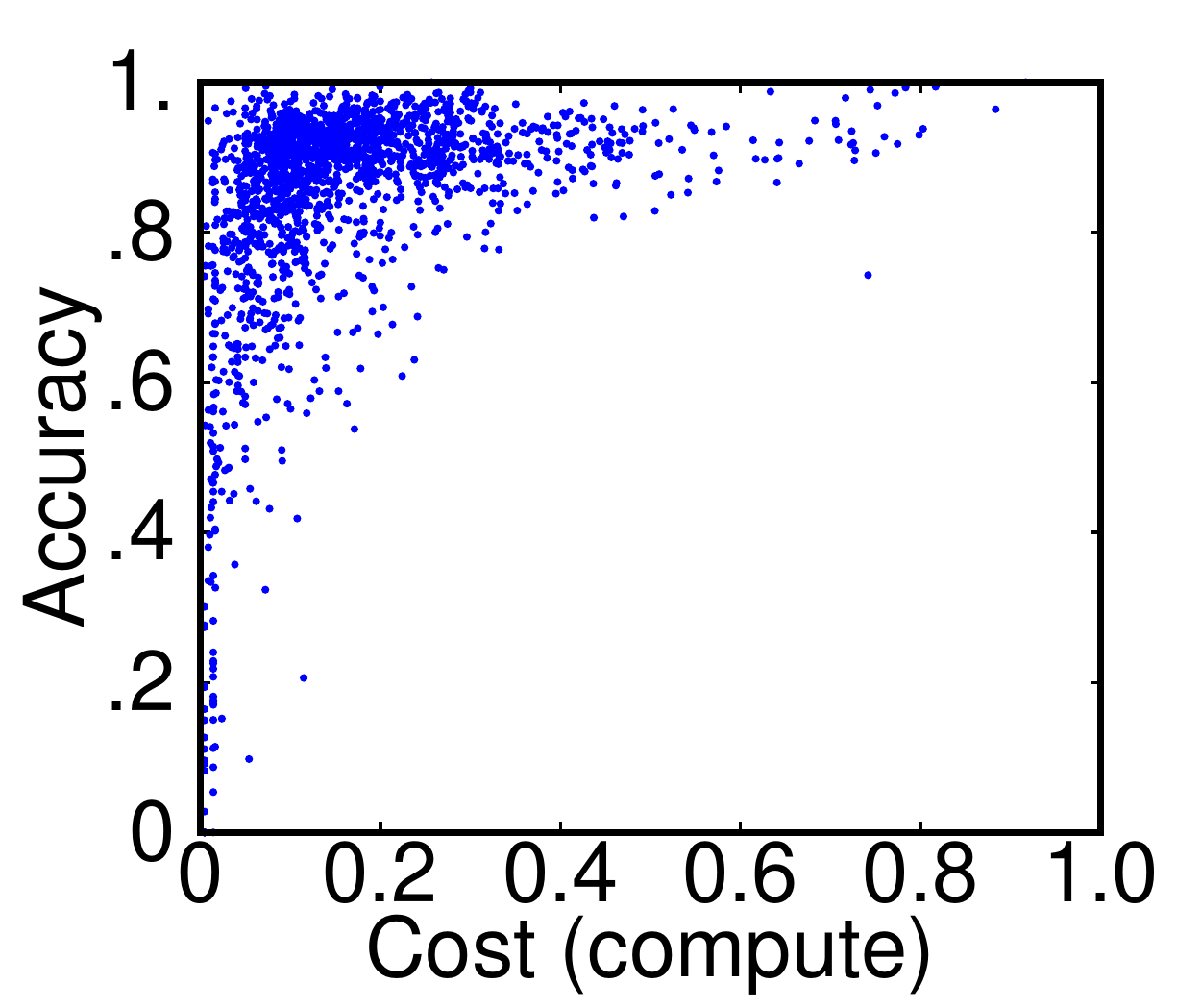}
    \vspace{-0.6cm}
    \subcaption{VideoStorm (Type 3)}
\end{subfigure}
\hspace{-0.42cm}
\vspace{-0.2cm}
\tightcaption{Significant performance variability across video segments. Each dot shows the accuracy and cost of a segment. 
}
\label{fig:variance-overall}
\end{figure*}

\mypara{Performance metrics used} We measure each \vp's performance using the following three metrics: 
\begin{packeditemize}
  \item {\bf Accuracy} is measured by the F1 score
of a \vp's detected objects~\cite{everingham2010pascal}. 
We obtain the ``ground truth'' results by running the full DNN on the uncompressed video frames (rather than the human-annotated labels). 
This way, any inaccuracy will be due to \vp designs (\eg video compression, DNN distillation), rather than errors made by the full DNN itself.
This is consistent with recent work (\eg~\cite{awstream,vigil,videostorm,chameleon,mullapudi2019online,noscope}). 
\item Normalized {\bf network cost} defines the data size sent by the camera to the server divided by the size of the original video. 
Reducing network cost is crucial when deploying \vps in bandwidth-constrained networks~\cite{noghabi2020emerging}.
\item Normalized {\bf compute cost} is the average GPU usage (on a NVidia GTX Titan Xp) per frame divided by that of the full DNN model.
Since the cost is normalized against running the full DNN model on the same GPU, it is less dependent on the particular choice of GPU.
Note that when a \vp (\eg Glimpse) reduces both compute and network costs, we will specify which is being considered.
\end{packeditemize}

We acknowledge that there are other aspects of \vp performance beyond these metrics.
Our choice of these metrics is based on two reasons.
First, these metrics are directly related to video content. 
For example, evaluating things like how adaptive a \vp is to bandwidth variations is important but 
deviates from our main goal of understanding the impact of video content.
Similarly, metrics like throughput, processing delay or energy consumption are crucial but also highly sensitive to the implementation details (\eg pipelining or parallelization) and hardware platform.
Second, these metrics can be translated into practical objectives. 
The feasibility of deploying a \vp depends on whether its costs fit the provisioned compute/network resources or the deployment budget.
\camready{Although we do not evaluate other performance metrics (\eg
  throughput, latency) explicitly, we believe they are highly
  correlated with the network and compute cost considered by our
  study.  For example, when a \vp reduces network cost by 2x, this
  saving can translate into serving 2x video streams while meeting the same
  inference accuracy target (i.e., 2x throughput). }

\tightsubsection{Key Findings}
\label{subsec:variability}


\begin{finding}
Performance of a \vp can vary dramatically even among videos of the same scenario.
\end{finding}



\noindent 
Following the traditional assumption (\S\ref{sec:background}), we test each \vp's performance (cost vs. accuracy) in one of the four {\em scenarios}: \{fix-positioned traffic monitoring cameras, moving dashboard cameras\} $\times$ \{on urban streets, or on highway\}.
Figure~\ref{fig:existing-coverage} summarizes each \vp performance range in each video (each over 20 minutes) in one ellipse. 
We see each \vp's performance can vary dramatically across videos in the {\em same} scenario.  
Such {\em performance heterogeneity} is prevalent across all 7 \vps and four scenarios considered by our study.




To reveal the full range of performance variability,
Figure~\ref{fig:variance-overall} plots the performance distributions
of the 5 \vps on all the video segments \htedit{in the coverage dataset} (each dot shows the performance on one segment).
While the overall trends align with findings of prior work (\vps trade accuracy drop for saving network/compute cost), we do see that each \vp has a significant performance variability across video segments.

\begin{table}[h]
\resizebox{0.48\textwidth}{!} {
\begin{tabular}{|c|c|c|c|c|c|c|}
\hline
\multirow{2}{*}{\textbf{\begin{tabular}[c]{@{}c@{}}Cost when acc.\\ is in {[}0.9, 0.95{]}\end{tabular}}} & \multicolumn{2}{c|}{\textbf{VAP type 1}} & \multicolumn{2}{c|}{\textbf{VAP type 2}} & \multicolumn{2}{c|}{\textbf{VAP type 3}} \\ \cline{2-7} 
 & AWStream & Glimpse & Vigil & NoScope & Glimpse & VideoStorm \\ \hline
Mean & 0.34 & 0.27 & 0.15 & 0.67 & 0.41 & 0.20 \\ \hline
Relative StdDev & 73\% & 64\% & 102\% & 48\% & 45\% & 68\% \\ \hline
\end{tabular}
}
\vspace{0.2cm}
\tightcaption{\camready{Even when we narrow the range of accuracy in Figure~\ref{fig:variance-overall} to [0.90,0.95],
the cost (network or compute) across segments could vary significantly. This can be seen from the relative standard deviation values in the table.}}
\label{table:variance}
\end{table}
Even when we restrict the accuracy to a small range ([0.90, 0.95]), the relative standard deviation of cost across segments can be 45-102\% and the gap between $5^{\textrm{th}}$ and $95^{\textrm{th}}$ percentiles is always over 90\% \camready{(shown in Table~\ref{table:variance}}). \camready{Here, relative standard deviation is defined as the ratio of the standard deviation to the mean, which
is a popular metric to measure the dispersion of a distribution.}

\ignore{
Notice that such variance is unlikely caused by tradeoffs between accuracy and cost, since accuracy is in a small range.
Similarly, when the cost is restricted to between 0.2 and 0.25, we still see a 49\% (AWStream), 95\% (NoScope), or 33\% (VideoStorm) difference between the $5^{\textrm{th}}$ and $95^{\textrm{th}}$ percentile accuracies.
}



\begin{finding}
Choice of optimal \vp is content-dependent.
\end{finding}

\begin{figure}[h]
\centering
\hspace{-1cm}
\begin{subfigure}[b]{0.178\textwidth}
    \includegraphics[width=\linewidth]{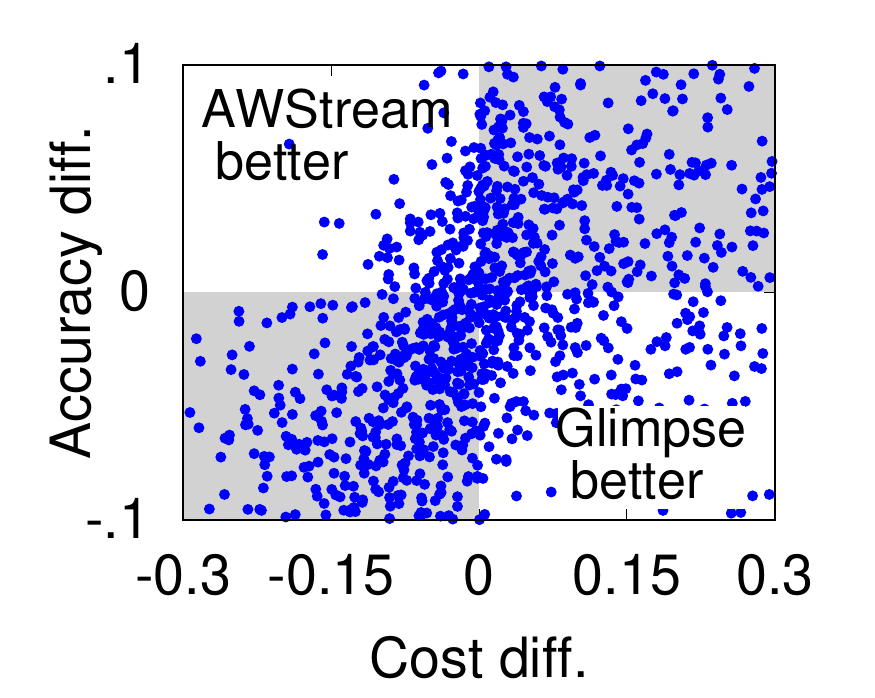}
    \vspace{-0.5cm}
    \subcaption{AWStream vs. \\Glimpse (Type 1)} 
\end{subfigure}
\hspace{-0.6cm}
\begin{subfigure}[b]{0.178\textwidth}
    \includegraphics[width=\linewidth]{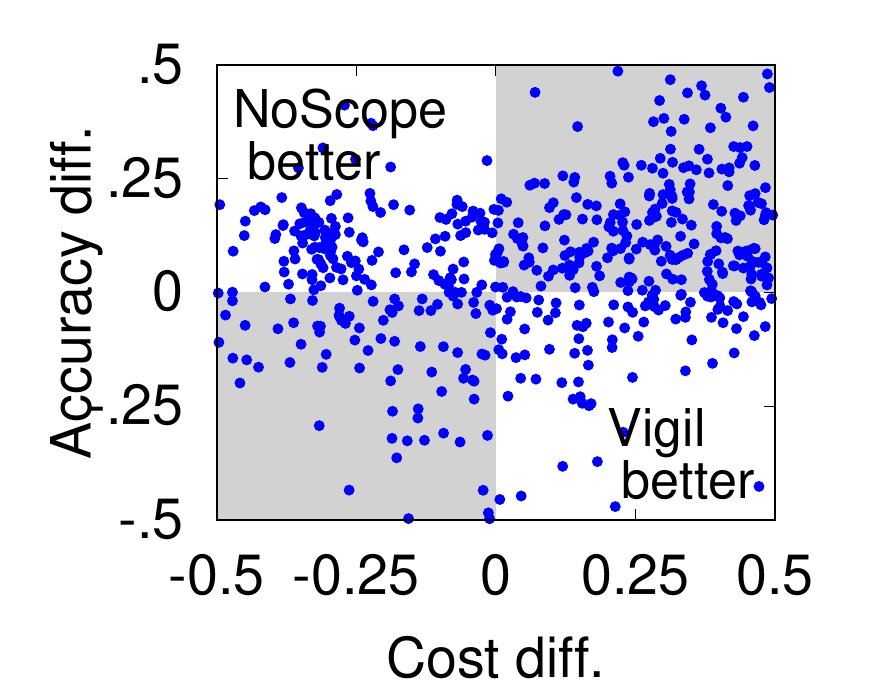}
    \vspace{-0.5cm}
    \subcaption{NoScope vs. \\Vigil (Type 2)}
\end{subfigure}
\hspace{-0.6cm}
\begin{subfigure}[b]{0.178\textwidth}
    \includegraphics[width=\linewidth]{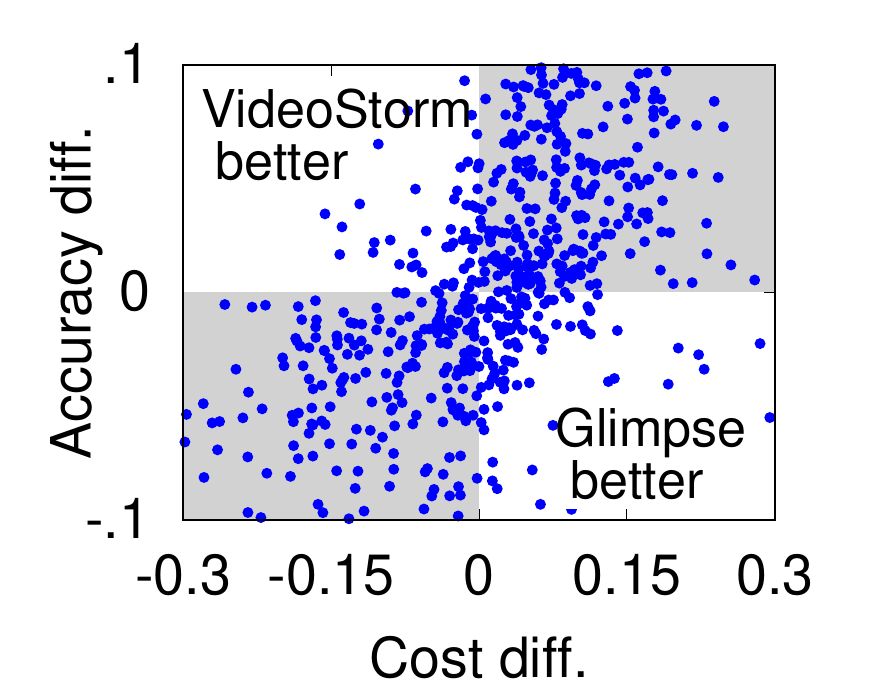}
    \vspace{-0.5cm}
    \subcaption{VideoStorm vs.\\ Glimpse (Type 3)}
\end{subfigure}
\hspace{-1cm}
\tightcaption{Each \vp outperforms the other on many video segments.
each dot shows the relative accuracy \& 
  cost between two \vps on one video segment. 
  }
\label{fig:order-spatial}
\end{figure}

\noindent 
Performance variance does not always lead to suboptimal choice of \vp,
if one \vp\ {\em always} outperforms others. 
 Unfortunately, that is not true for \vps. 
\htedit{We illustrate this by comparing \vps in pairs. 
In each pair, one \vp acts as a ``reference'', and we subtract the other \vp's cost
and accuracy on each video segment by those of the reference. 
Figure~\ref{fig:order-spatial} shows
the results of three \vp pairs and marks the region where one \vp is
strictly better than the other (higher accuracy {\em and}
lower cost). }  Clearly, the choice of best \vp varies
across video segments and is content-dependent. Thus, it is crucial for \vp operators and developers to understand under {\em what videos} would one \vp perform better than others.

Together, these findings cast doubt over the current \vp evaluation
methodology:

\begin{statement}
  \htedit{{\bf Key Takeaway:} Empirically testing and comparing \vps on
   some specific video workloads can be incomplete.}
\end{statement}

\tightsubsection{Discussion}

Our measurement has shown that if a \vp is evaluated on only a handful
of videos, the results may 
fail to reveal its true performance range and variance in a target
scenario. An immediate response is ``why not using a better test
dataset?'' 

\myparaq{Why not using a representative dataset} Intuitively, with a set of ``representative'' videos per scenario, we can get
the most common \vp performance by testing \vps on these videos.
Unfortunately, this solution is impractical for two reasons. 
First, cameras deployed at different locations or future locations will likely generate video workloads with different content characteristics beyond those captured by the empirical tests.
Second, since video analytics applications are continuously evolving, representative workloads do not yet exist. 
Thus, these tests might overestimate/underestimate the
\vp performance  and \htedit{lead to wrong choice of \vp in  deployment}.

\myparaq{Why not using a larger dataset} Testing a \vp on a
  larger number of videos might offer a more complete view of its
  performance range and variance.  Yet a ``just adding data'' approach will provide little
  insight on performance distribution on videos outside of the
  test dataset,  and why a \vp's performance varies across
  videos.


\tightsection{\hspace{-0.1cm}Achieving Performance Clarity}
\label{sec:clarity}
Different from prior work that evaluates \vps using only empirical tests, \htedit{we propose a new methodology for \vp evaluation: {\bf achieving performance clarity}. The goal of performance clarity is to not only identify a \vp's performance under a wide range of video content, but also characterize how video content characteristics affect its performance.  This produces a comprehensive and transparent assessment of VAP performance.  } In the following, we first present the key concept behind performance clarity and its benefits, and then discuss potential solutions to achieve performance clarity. 

\camready{To facilitate the discussion below, Table~\ref{tab:terminologies} summarizes the key terminologies and notations used by our work.}

%

{\small 
\begin{table}[]
{\color{black}\begin{tabular}{cc}
\hline
\textbf{Terminology} & \textbf{Definition} \\ \hline
\begin{tabular}[c]{@{}r@{}}Video content \\ features \end{tabular} & \begin{tabular}[c]{@{}l@{}}Features that measure content-level characteristics \\ of a video (\eg avg object speed). See \S\ref{subsec:features}.\end{tabular} \\ \hline
\begin{tabular}[c]{@{}r@{}}Performance \\ Clarity (PC)\end{tabular} & \begin{tabular}[c]{@{}l@{}}Comprehensive performance assessment of VAPs \\ under different video content features. See \S\ref{subsec:content-correlation}.\end{tabular} \\ \hline
\begin{tabular}[c]{@{}r@{}}PC Profile\\ ($\Perf_{\Pipeline}$)\end{tabular} & \begin{tabular}[c]{@{}l@{}}A lookup table that maps video content features to \\ performance of VAP $\Pipeline$ (\eg Figure~\ref{fig:workflow-profiler})\end{tabular} \\ \hline
\begin{tabular}[c]{@{}r@{}}Cost-saving \\ Strategy\end{tabular} & \begin{tabular}[c]{@{}l@{}}A particular heuristic to save computer/network \\ cost. See \S\ref{subsec:primitives}.\end{tabular} \\ \hline
\begin{tabular}[c]{@{}r@{}}VAP \\ Primitives\end{tabular} & \begin{tabular}[c]{@{}l@{}}A set of cost-saving strategies that seek to reduce \\ same type of redundancies. See \S\ref{subsec:primitives}.\end{tabular} \\ \hline
\end{tabular}
}
\tightcaption{\camready{Definition of terminologies used in \name.}}
\label{tab:terminologies}
\end{table}
}


\tightsubsection{Defining Performance Clarity (\Pc)}
\label{subsec:content-correlation}

\begin{figure}[t]
\includegraphics[width=0.41\textwidth]{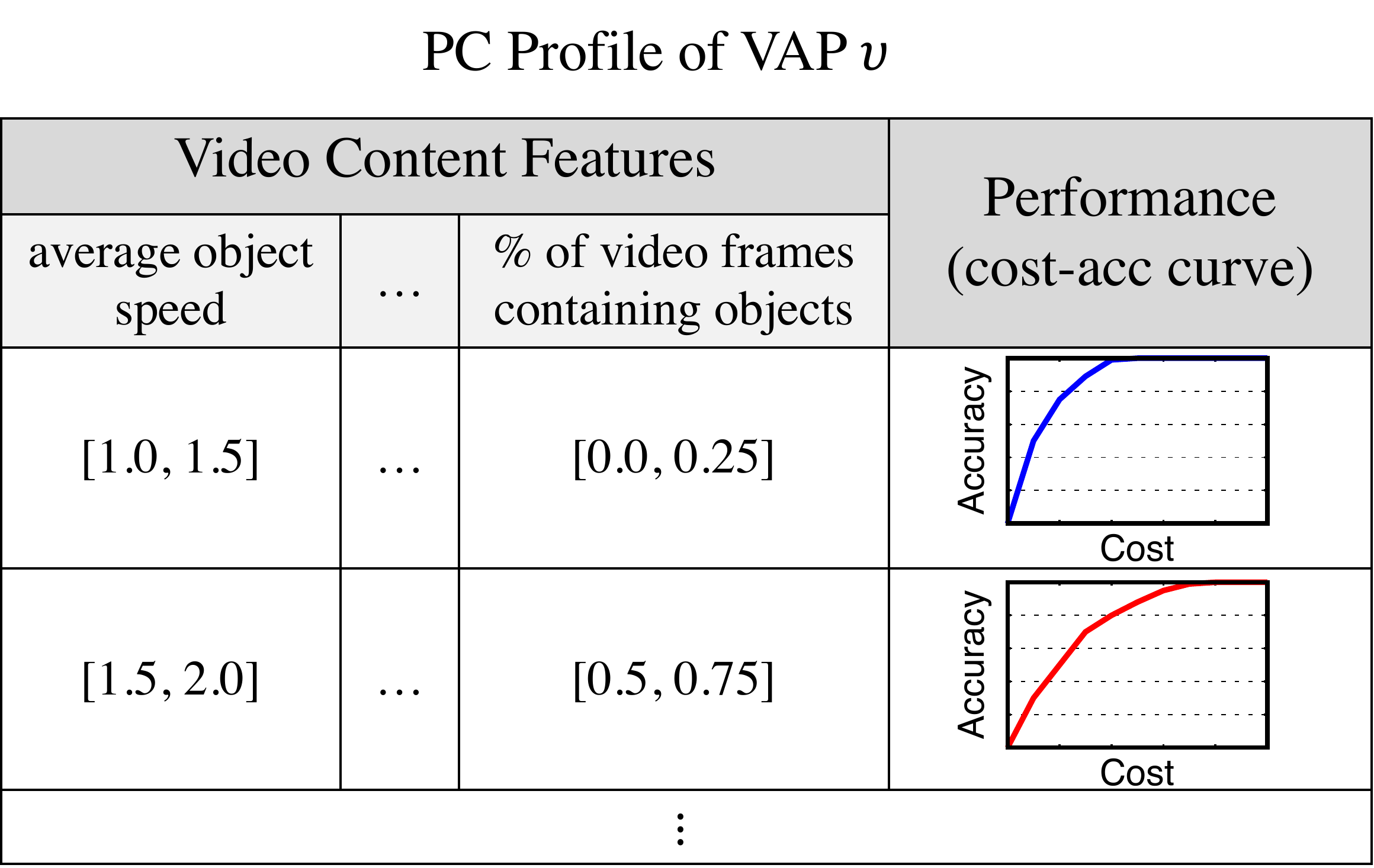}
\tightcaption{\camready{An abstract illustration of a \vp $v$'s performance
    clarity (PC) profile. 
    Compared to either testing a \vp on few videos or reporting its performance distribution over many videos, this profile provides a more complete picture of the VAP's performance by describing its relationship with video content features, which drastically reduces the ambiguity of performance compared to those in Figure~\ref{fig:variance-overall}.
    } 
}
\label{fig:workflow-profiler}
\end{figure}

The {\em performance clarity}~(\pc) of a \vp defines {\em how video content features affect the \vp's performance}\footnote{While performance clarity reveals correlations between content features and \vp performance, it does not equal to interpretation of DNNs or \vps.}.
Formally, \pc of a \vp $\Pipeline$ is a lookup table $\Perf_{\Pipeline}$ that maps from a point $\FeatureValVec$ in the space of video content features to $\Pipeline$'s performance (in cost and accuracy) on videos that match $\FeatureValVec$. \htedit{This is illustrated by  Figure~\ref{fig:workflow-profiler}. Compared to existing evaluations that are either incomplete (\eg single-scenario tests in Figure~\ref{fig:existing-coverage}) and/or ambiguous (\eg high performance variability in Figure~\ref{fig:variance-overall}), \pc offers a comprehensive and clear characterization of \vp performance and its variation.}

\begin{figure}[t]
\hspace{-0.2cm}
\begin{subfigure}{.127\textwidth}
    \centering
    \includegraphics[width=\linewidth]{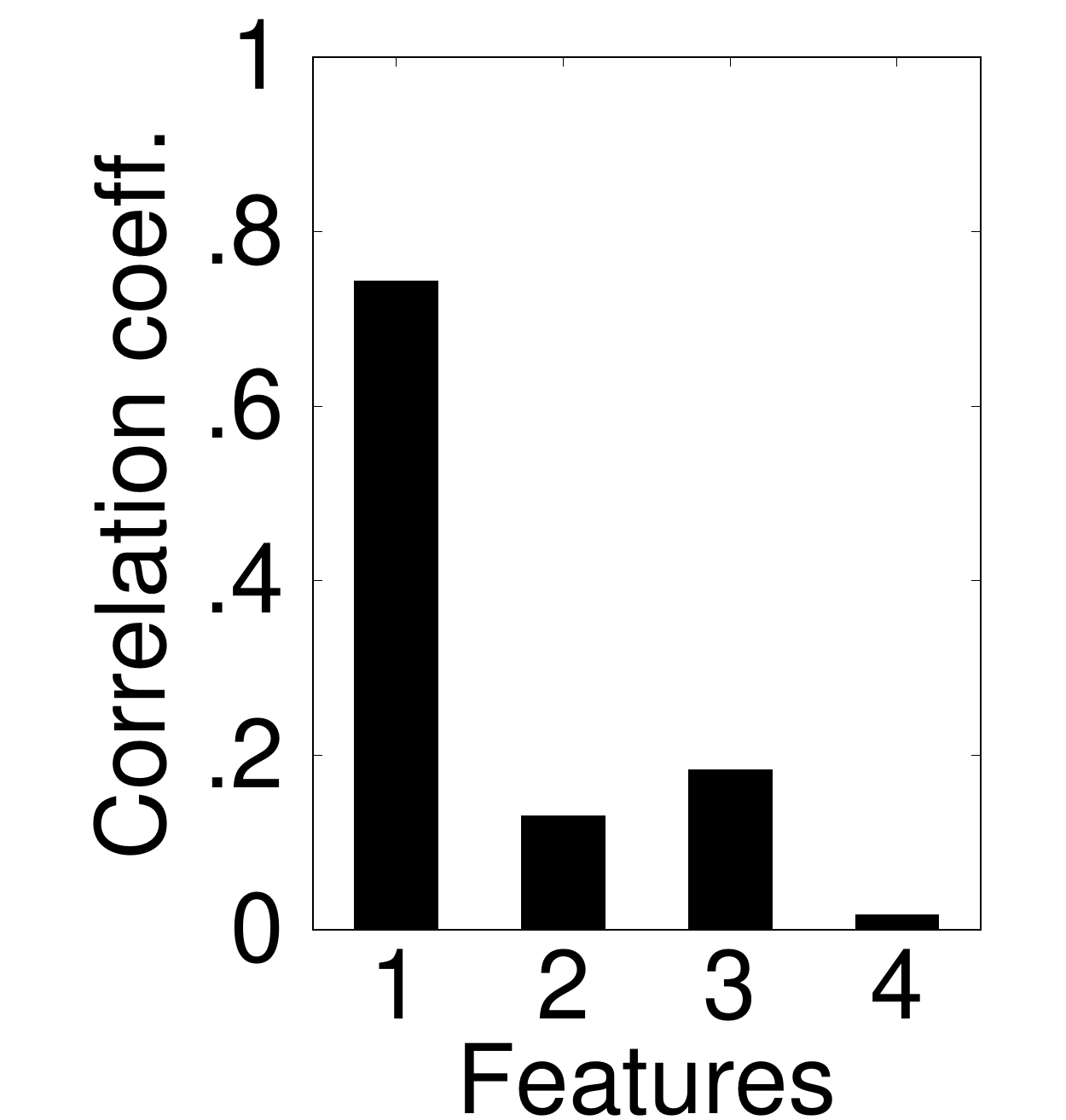}  
    \vspace{-0.45cm} 
    \caption{VideoStorm}
    \label{fig:feature-correlation-videostorm}
\end{subfigure}
\hspace{-0.3cm}
\begin{subfigure}{.127\textwidth}
    \centering
    \includegraphics[width=\linewidth]{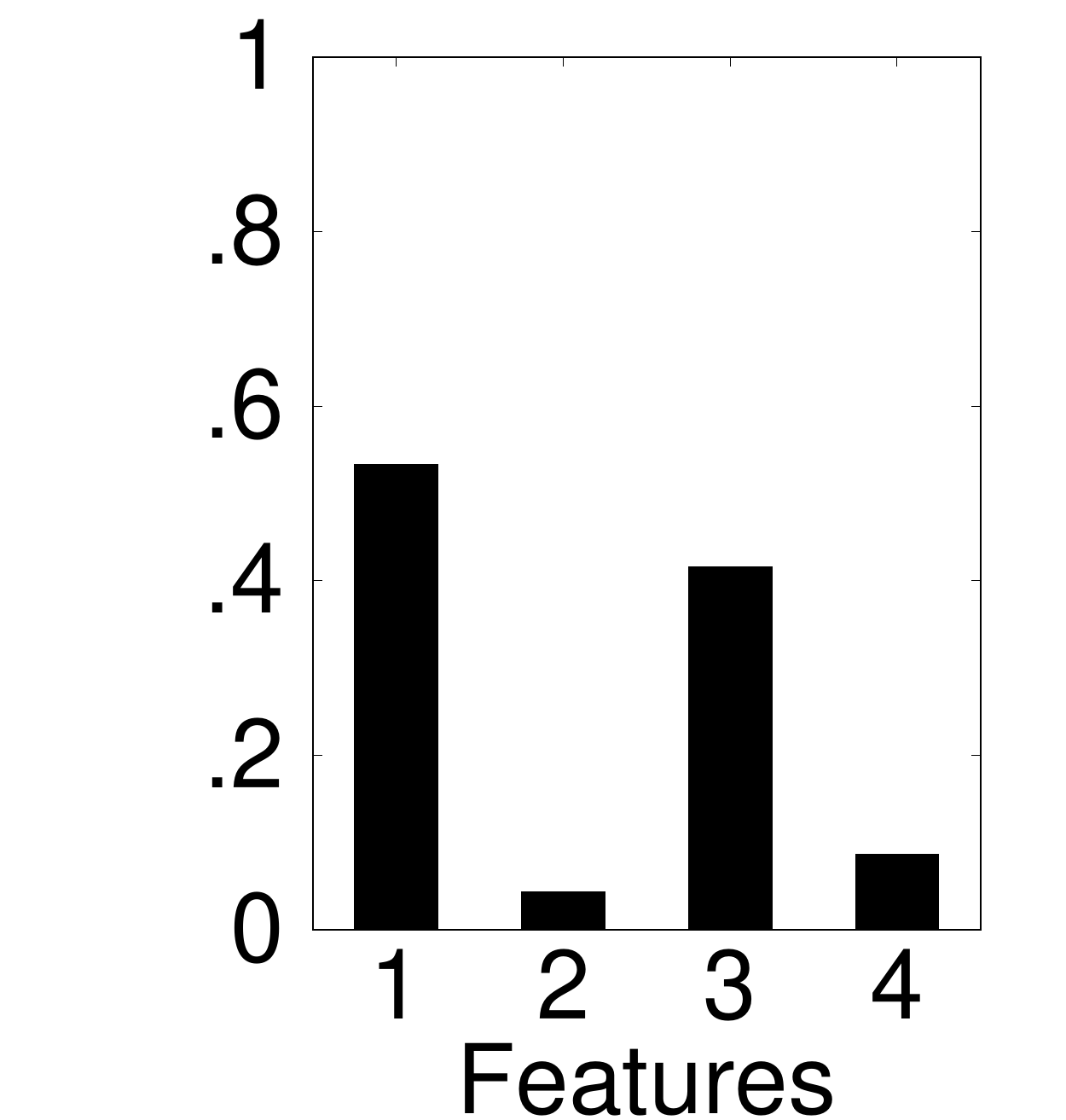}  
    \vspace{-0.45cm} 
    \caption{AWStream}
    \label{fig:feature-correlation-awstream}
\end{subfigure}
\hspace{-0.3cm}
\begin{subfigure}{.127\textwidth}
    \centering
    \includegraphics[width=\linewidth]{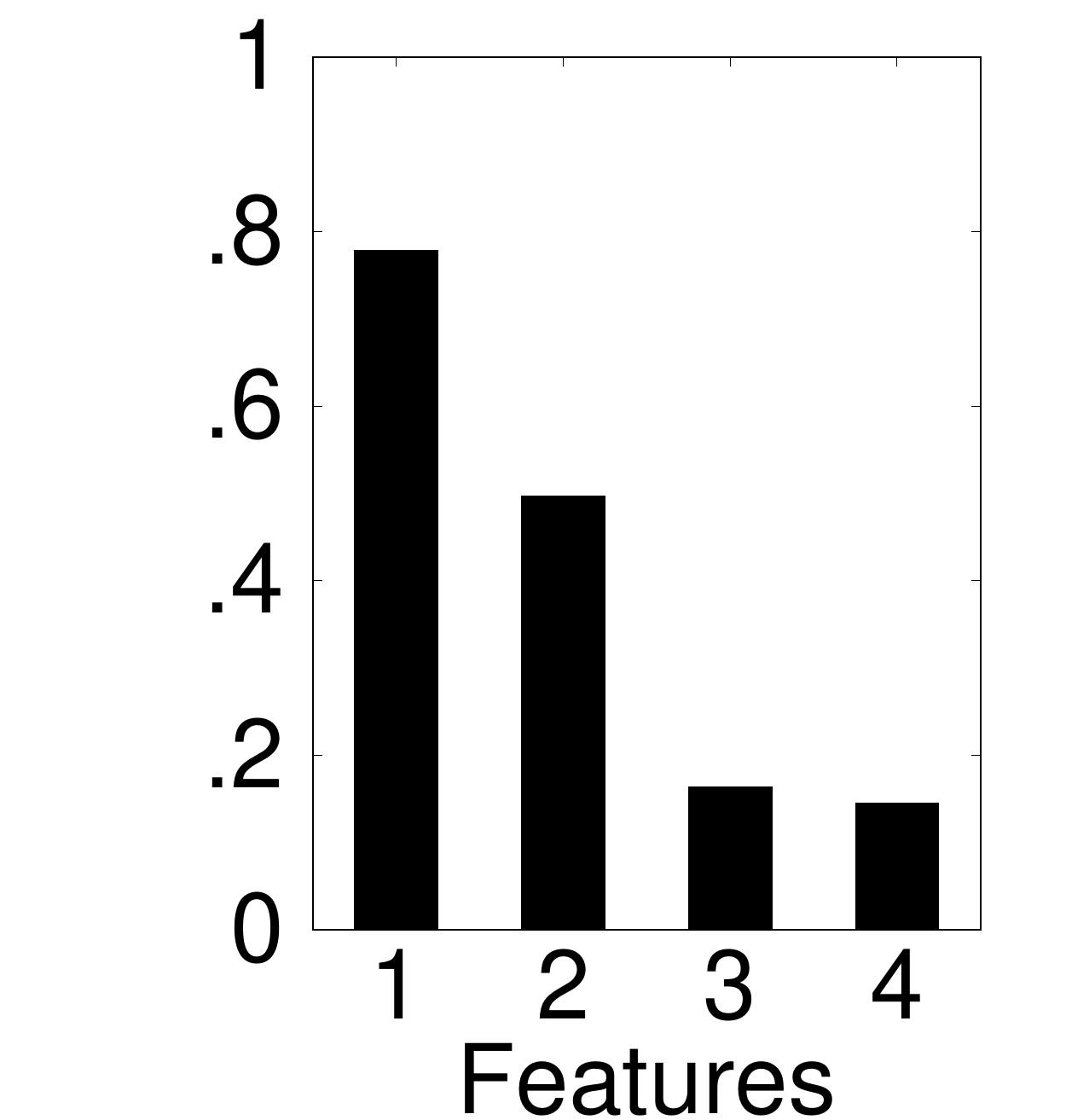} 
    \vspace{-0.45cm} 
    \caption{Glimpse}
    \label{fig:feature-correlation-glimpse}
\end{subfigure}
\hspace{-0.3cm}
\begin{subfigure}{.127\textwidth}
    \centering
    \includegraphics[width=\linewidth]{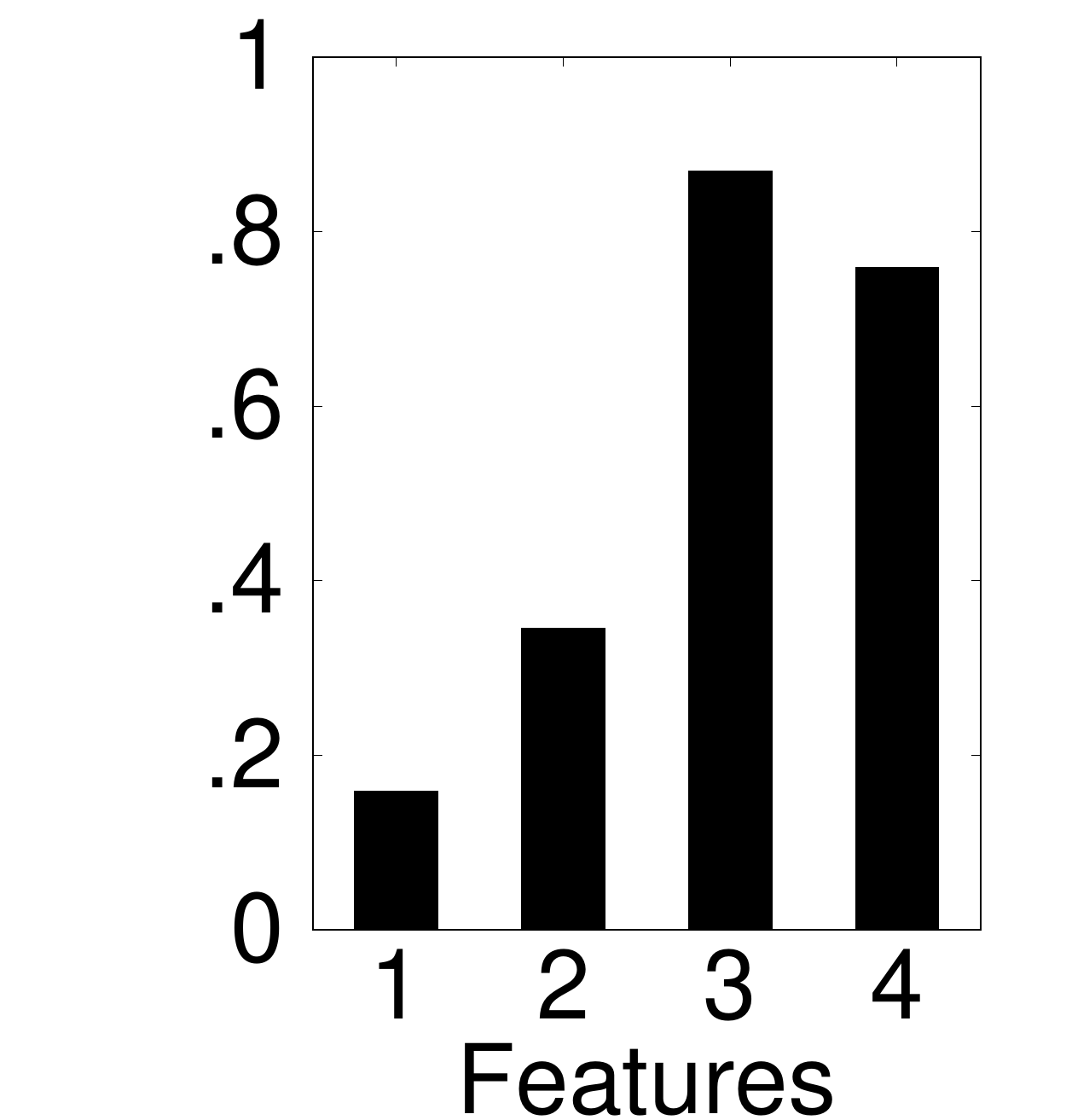}  
    \vspace{-0.45cm} 
    \caption{NoScope}
    \label{fig:feature-correlation-noscope}
\end{subfigure}
\hspace{-0.2cm}
\vspace{-0.1cm}
\tightcaption{\htedit{Strong correlation between video content features and \vp performance. The four features are: ($\Feature_1$) \feature{avg. object speed}, ($\Feature_2$) \feature{\% of frames with objects}, ($\Feature_3$) \feature{10\%ile of per-object area}, and ($\Feature_4$) \feature{avg. confidence score per object}.}
\vspace{-0.1cm}
}
\label{fig:relations-features}
\end{figure}

The key insight behind \pc is the following.
It is the {\em \vp performance's dependencies on video content features} that cause the \vp performance variabilities.
As these content features vary across videos (in the same scenario), so does \vp performance. \htedit{To illustrate this, we featurize each video in our \htedit{coverage} dataset along four content features (more features discussed later in \S\ref{subsec:features}), and plot in Figure~\ref{fig:relations-features}} the Pearson's correlation coefficients between individual features and cost of \vps when keeping accuracy between 0.9 and 0.95 (to avoid cost variance caused by accuracy variance).
We single out the impact of each feature by restricting other features to a small range less than 50\% of their respective value ranges.
The results show a strong correlation between each \vp's performance and the content features.

\mypara{Benefits of \pc} \htedit{A \vp $v$'s performance variation and its content dependency come from the $v$'s design, {\em i.e.\/} they are {\em inherent} to $v$.  Thus  $v$'s \pc profile ($\Perf_{\Pipeline}$) can offer useful insights on its design and deployment. Below are two usage cases. }

\begin{packedenumerate}
\item To {\bf estimate $v$'s performance on any target video}, we can directly combine $\Perf_{\Pipeline}$ with the content feature distribution of the video, which can be quickly obtained by scanning through the video. \htedit{The computation cost is significantly less than running $v$ on the video (verified in \S\ref{sec:usecase}).}

\item To {\bf identify when one \vp outperforms another}, we can directly compare two \vps' \pc profiles to identify in which parts of the content feature space is one \vp  better.  Again there is no need to run \vps on any video. 
\end{packedenumerate}
\htedit{
  Later in \S\ref{sec:usecase} we use these two tasks to evaluate the accuracy and benefits of our \pc profiler \name.}

\tightsubsection{Feature-based Profiling: Why It Fails} 


Building an accurate \pc profile is challenging.  A straightforward solution is to create a corpus of videos that span all combinations of \htedit{relevant}  content feature values, and test \vps on these videos.  Unfortunately, this can be prohibitively expensive due to the complex relationship between \vp performance and content features.  Specifically, our measurement study ({\em e.g.\/} Figure~\ref{fig:relations-features}) lead to  two observations. 
\begin{packeditemize}
\item{\em Heterogeneity impact of features:} 
Different {\vps} are affected by {\em different} sets of features.
For example, VideoStorm is sensitive to average object speed ($\Feature_1$) but not per-object area ($\Feature_3$); yet NoScope is highly sensitive to $\Feature_3$ but not $\Feature_1$.

\item {\em Combinatorial impact of features:}
A \vp can be affected by {\em multiple} features. 
For instance, Glimpse is highly correlated with the features of object speed ($\Feature_1$) and fraction of frames with objects ($\Feature_2$), and AWStream is sensitive to $\Feature_1$ and $\Feature_3$.
Therefore, {\em it is insufficient to test \vps on videos that vary along only one feature at a time}.
\end{packeditemize}
Thus, to cover all possible feature value combinations, we need 
\htedit{$O(n^{|\FeatureVec|})$} videos, where $\FeatureVec$ is the list of content features and $n$ is the number of possible values per feature.
To put it into perspective, let us assume that there are 7 content features, each having 4 distinct value buckets (\eg low, median, high and very high), and we need three 30-second videos \htedit{to measure \vp performance} for each of the $4^7$ feature value combinations.
These are not overestimation: there are at least 7 content features that might affect DNN accuracy or \vp performance (see \S\ref{subsec:features}), and in our dataset we split each feature in four buckets as well.
The resulting dataset would be over {\bf 400 hours}, much longer than any \vp test datasets ever created.
Since  many \vps do not reduce compute cost, evaluating their performance on this hypothetical dataset would take 400 hours even when using one NVidia GTX Titan X GPU card running the state-of-the-art object detector at 30fps~\cite{google-benchmark}.

\tightsubsection{Proposed: Primitive-based Profiling}

Instead of profiling a \vp as a monolithic entity, 
we modularize it into multiple primitives (\S\ref{subsec:primitives}), each of which can be profiled separately.
The rationale is two-fold.
\begin{packedenumerate}
\item {\bf A primitive is affected by fewer features than a \vp.}
Each primitive only leverages, and is thus affected by, a particular \htedit{set of video content characteristics}. 
For example, many \vps reduce video frame rates to save cost, and its efficacy depends only on temporal-related features like object speeds.
Yet these features have little impact on ``orthogonal'' techniques like image downsizing or model compression.

\item {\bf Primitives have independent impacts on a \vp's performance.}
As we will show in \S\ref{subsec:primitives}, the performance of a \vp can be approximated by multiplying the performance of each primitive when other primitives are set to their corresponding most accurate, expensive strategies.
In other words, these primitives can be profiled individually, based on which the full \vp performance can be \htedit{constructed}. 
\end{packedenumerate}


\mypara{Reducing profiling cost} Since each primitive is profiled using {\em only} the video features relevant to its cost-saving strategy,  the \vp profiling overhead can be drastically reduced, 
from $O(n^{|\FeatureVec|})$ to $O(n^{|\FeatureVec_1|}+n^{|\FeatureVec_2|}+\cdots)=O(n^{\textrm{max}_i|\FeatureVec_i|})\ll O(n^{|\FeatureVec|})$, where 
$\FeatureVec_i$ is the feature set related to the $i^{\textrm{th}}$ primitive.  Using primitive-based profiling, our eventual dataset consists of only {\bf 67.5 minutes} of videos, \htedit{more than two} orders of magnitudes less than that of feature-based profiling (400 hours)!


\tightsection{Yoda: Practical VAP Profiling}
\label{sec:design}

We now describe our design of \name, the first \vp benchmark to achieve performance clarity. \name builds a \pc profile for each \vp, by applying the  aforementioned primitive-based profiling.  In the following,  we first present how \name modularizes a \vp  into independent primitives~(\S\ref{subsec:primitives}) and chooses content features and benchmark videos to profile each primitive~(\S\ref{subsec:features}), followed by  two core functions offered by \name: \vp profiler and \vp performance predictor~(\S\ref{subsec:components}).

\subsection{Modularizing \Vps into Primitives}
\label{subsec:primitives}

A \vp may employ one or more {\em cost-saving strategies} to reduce redundancies in video frames, pixels, and DNN parameters. \htedit{Observing this inherent modularity}, \name categorizes these strategies into three {\em primitives} (see Table~\ref{tab:summary}).\footnote{Some prior work also reduces redundancies across multiple concurrent queries~\cite{mainstream} or camera streams~\cite{rexcam-hotmobile}. We leave them to future work.}

\begin{table}[t]
\resizebox{0.49\textwidth}{!}
{
\begin{tabular}{c >{\centering\arraybackslash}m{2.2cm}>{\centering\arraybackslash}m{1.95cm} >{\centering\arraybackslash}m{1.95cm}}
 \hline
  {\em \Vp} & {\em Temporal pruning} &  {\em Spatial pruning } & {\em Model pruning} \\ \hline\hline
VideoStorm{\tiny\cite{videostorm}}       & \ding{52} {\tiny(uniform sampling)}                                                  & \ding{52}{\tiny(quality downsize)}                                               & \ding{52} {\tiny(model selection)}                                                            \\ \hline
NoScope{\tiny\cite{noscope}}                 & \ding{52} {\tiny(diff-triggered)}                                           &                                                           & \ding{52}{\tiny(specialization)}                                                           \\ \hline
AWStream{\tiny\cite{awstream}}          & \ding{52}   {\tiny(uniform sampling)}                                                 & \ding{52}{\tiny(quality downsize)}                                               &                                                                                         \\ \hline
Glimpse{\tiny\cite{glimpse}}        & \ding{52} {\tiny(diff-triggered)}                                                  &                                                           & \ding{52}{\tiny(fixed tiny model)}   \\ \hline
Vigil{\tiny\cite{vigil}}            & \ding{52} {\tiny(diff-triggered)}                                                & \ding{52}  {\tiny(region cropping)}                                                &                                                                                         \\ \hline
Chameleon{\tiny\cite{chameleon}}           & \ding{52} {\tiny(uniform sampling)}                                                    & \ding{52}{\tiny(quality downsize)}                                               & \ding{52}{\tiny(model selection)}                                                           \\ \hline
VideoEdge{\tiny\cite{videoedge}}              & \ding{52}{\tiny(uniform sampling)}                                                & \ding{52}{\tiny(quality downsize)}                                                &                                                                                         \\ \hline
DDS{\tiny\cite{dds-hotcloud}}                     &                                                            & \ding{52}  {\tiny(region cropping)}                                               &                                                                                \\ \hline
EAAR{\tiny\cite{eaar}}                     &         \ding{52} {\tiny(diff-triggered)}         & \ding{52}  {\tiny(region cropping)}                                               &                                                                                \\ \hline
Reducto{\tiny\cite{reducto}}        & \ding{52} {\tiny(diff-triggered)}                                                  &                                                           &
  \\ \hline
WEG{\tiny\cite{fast}}         &                                                            &                                                           &  \ding{52}{\tiny(specialization)}                                                                                \\ \hline
\end{tabular}
}
\vspace{0.cm}
\tightcaption{\htedit{Modularizing some example {\vps} into primitives.}
\vspace{-0.5cm} }
\label{tab:summary}
\end{table}

\begin{packeditemize}
\item {\bf Primitive \#1: Temporal pruning} drops frames to reduce inter-frame redundancies using at least two strategies.
{\em Uniform frame selection} (\eg~\cite{awstream,videostorm}) uniformly samples a fraction of frames for further analysis and then carries over their detected objects to future unsampled frames (\eg via object tracking).
It works well if neighboring frames are similar.
{\em Trigger-based frame selection} (\eg~\cite{noscope,glimpse}) skips frames until a heuristic (\eg significant difference between frames) signals potential arrivals of new objects.
It works well when most frames have few objects of interest.

\item {\bf Primitive \#2: Spatial pruning} reencodes video to reduce redundancies among pixels. Specifically,
{\em image quality downsizing} (\eg~\cite{awstream,chameleon}) reduces the video quality (\eg from 1080p to 360p), which still achieves high accuracy if objects are large. Another strategy, {\em region cropping} (\eg~\cite{vigil,dds-hotcloud}), saves bandwidth by encoding only pixels relevant to the task.
It can be very effective in, for instance, traffic videos where most vehicles/pedestrians appear small.

\item{\bf Primitive \#3: Model pruning} leverages the fact that videos often have specific object classes/scenes (\eg traffic videos contain mostly vehicles/pedestrians with static background), \htedit{ and trims the full DNN to reduce compute cost while still achieving high accuracy.}
{\em Model selection} (\eg~\cite{videostorm,chameleon}) picks a simple yet accurate DNN model from a few pre-trained models with various capacities.
{\em Model specialization} (\eg~\cite{noscope,fast}) trains a smaller DNN just for particular scenes/objects and if it fails, falls back to the full DNN.

\end{packeditemize}

\noindent Finally, for each primitive, \name also defines an {\bf oracle strategy} that does no cost reduction:
100\% frame selection (for temporal pruning,
original video quality (for spatial pruning), and
full-size DNN (for model pruning).
Since the primitives essentially trade accuracy for cost savings, these oracle strategies serve as
the most accurate yet most costly strategies.

\begin{figure*}[t]
  \includegraphics[width=1\linewidth]{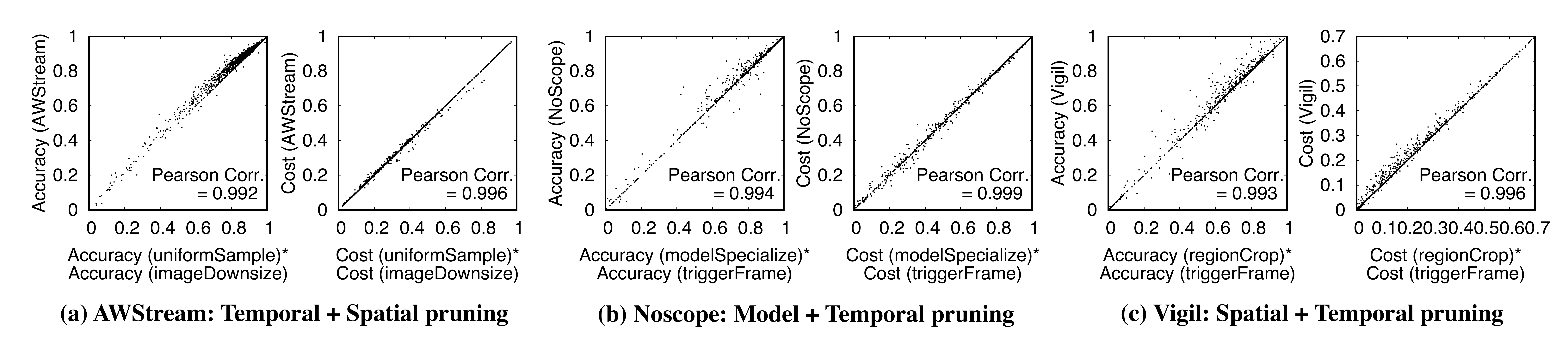}
  \vspace{-0.7cm}
\tightcaption{Empirical validation of the cross-primitive independence on AWStream, NoScope and Vigil: \vp performance can be approximated by the product of individual cost-saving strategies' performance. Each dot is a video segment in our dataset. Table~\ref{tab:independence-correlation} validates the independence property on more strategy pairs.
}
\label{fig:deconstruction-performance}
\end{figure*}

\begin{table*}[t!]
{\footnotesize
\begin{tabular}{|c|c|c|c|c|c|c|c|c|c|c|c|c|}
\hline
 & \multicolumn{4}{c|}{\textbf{Model pruning + Temporal pruning}} & \multicolumn{4}{c|}{\textbf{Spatial pruning + Temporal pruning}} & \multicolumn{4}{c|}{\textbf{Spatial pruning + Model pruning}} \\ \cline{2-13}
 & $M_1+T_1$ & $M_2+T_1$ & $M_1+T_2$ & $M_2+T_2$ & $S_1+T_1$ & $S_2+T_1$ & $S_1+T_2$ & $S_2+T_2$ & $S_1+M_1$ & $S_2+M_1$ & $S_1+M_2$ & $S_2+M_2$  \\ \hline
\textbf{Accuracy} & 0.994 & 0.997 & 0.991 & 0.994 & 0.992 & 0.993 & 0.989 & 0.993 & 0.955 & 0.914 & 0.993 & 0.999 \\ \hline
\textbf{Cost} & 1 & 1 & 1 & 0.999 & 0.996 & 0.987 & 0.999 & 0.996 & 1 & 1 & 1 & 1 \\ \hline
\end{tabular}
}
\vspace{0.3cm}
\tightcaption{\editsec{Independence property between any pair of strategies from model-pruning strategies ($M_1$: model selection, $M_2$: model specialization), temporal-pruning strategies ($T_1$: uniform sampling, $T_2$: trigger-based frame selection), and spatial-pruning strategies ($S_1$: image downsizing, $S_2$: region cropping). Each value shows the Pearson's correlation coefficient between the performance (accuracy or network cost) when the two strategies are combined and the product of the performance when each strategy is used separately.
The high correlations suggest the cross-primitive independence is common.
\vspace{-0.3cm}}}
\label{tab:independence-correlation}
\end{table*}

\mypara{Independence across primitives}
As different primitives seek to remove {\em agnostic} redundancies in video/model, we empirically observe that individual primitives affect \vp performance independently.
For instance, the efficacy of spatial-pruning strategies
is largely dependent on object sizes/shapes,
whereas the efficacy of model-pruning strategies
depends on the scene complexity or skewness in object class distributions, both of which are agnostic to object sizes/shapes.

\htedit{Figure~\ref{fig:deconstruction-performance} and Table~\ref{tab:independence-correlation} empirically validate the property of cross-primitive independence on existing \vps.}
For a \vp $v$, we first measure performance of each individual strategy by replacing other strategies with their respective oracle strategies.
For instance, we measure the performance (cost and accuracy) of $v$'s spatial-pruning strategy by running it on all video segments in the coverage set while setting $v$'s temporal-pruning primitive to the oracle strategy (full frame rate).
We then compare the performance of the full \vp and the {\em multiplication} of performance of its individual primitives.
We do so using Pearson's correlation.
\editsec{
Using this methodology, Table~\ref{tab:independence-correlation} shows that the independence property largely holds on different pairs of strategies from two distinct primitives.
Figure~\ref{fig:deconstruction-performance} shows three concrete examples (AWStream, NoScope and Vigil), where each \vp's performance (both accuracy and cost) closely matches the multiplication of its primitives.
}

\editsec{We acknowledge that the cross-primitive independence is empirical and there can be exceptions to it.
For instance, when spatial pruning downsizes video frames to an extremely low resolution, no object can be detected regardless of the temporal pruning strategy.
In this case, the efficacy of temporal pruning is affected by spatial pruning, though this is unlikely to occur in practice as {\vps}  aim to maintain a high accuracy.

Nevertheless, we believe cross-primitive independence property is still valuable.
By breaking down each \vp to individual primitives (strategies) each related to a subset of content characteristics, we can dramatically reduce the cost of profiling \vps in an exponential feature space.
Likewise, developers of new strategies can apply the same method (of Figure~\ref{fig:deconstruction-performance}) to verify if the independence assumption holds.}

\tightsubsection{\hspace{-0.1cm}Selecting Benchmark Features and Videos}
\label{subsec:features}
\htedit{Following the above discussion, \name profiles a \vp by first profiling its individual primitives and assembling them to construct the full \vp profile.  To profile a primitive, \name first selects its associated video features and video datasets.}


\begin{table}[t]
\color{black}{
\resizebox{0.475\textwidth}{!}
{
\begin{tabular}{l|l|l}
\hline
\multicolumn{2}{c|}{\textbf{Video content feature}} & \multicolumn{1}{c}{\textbf{Definition}} \\ \hline \hline
\multirow{3}{*}{Per object} & \feature{object speed} & \begin{tabular}[c]{@{}l@{}}the reciprocal of IoU between the \\ bounding boxes of the same object \\ detected in two consecutive frames\end{tabular} \\ \cline{2-3} 
 & \feature{object area} & \begin{tabular}[c]{@{}l@{}}the bounding box size of each object \\ divided by the frame size\end{tabular} \\ \cline{2-3} 
 & \begin{tabular}[c]{@{}l@{}} \feature{confidence} \\ \feature{score}\end{tabular} & \begin{tabular}[c]{@{}l@{}}the confidence score of each detected \\ object given by the full DNN\end{tabular} \\ \hline
\multirow{2}{*}{Per frame} & \begin{tabular}[c]{@{}l@{}}\feature{total area of} \\ \feature{objects}\end{tabular} & \begin{tabular}[c]{@{}l@{}}fraction of pixels covered by all object\\ bounding boxes in a frame\end{tabular} \\ \cline{2-3} 
 & \feature{object count} & the number of objects per frame \\ \hline
Per second & \begin{tabular}[c]{@{}l@{}}\feature{object arrival} \\ \feature{rate}\end{tabular} & \# of new arrival objects per second \\ \hline
Per segment & \begin{tabular}[c]{@{}l@{}}\feature{\% frames} \\ \feature{with objects}\end{tabular} & \begin{tabular}[c]{@{}l@{}}percentage of frames containing \\ objects\end{tabular} \\ \hline
\end{tabular}
}
}
\vspace{-.1cm}
\tightcaption{\camready{Summary of video content features} \vspace{-.5cm}}
\label{tab:features}
\end{table}

\mypara{Feature selection}  \htedit{We first create a set of 43 {\em candidate} content features}
\jceditnew{based on 7 general content-level features \camready{(summarized in Table~\ref{tab:features})} known in the computer vision community to influence object detection accuracy (\eg~\cite{coco,everingham2010pascal}) and potentially \vp performance.}
\camready{Among them, 6 features are defined either per object, per frame, or per second.}
We pair them with 7 statistics per video segment: mean, standard deviation, and  \{10,~25,~50,~75,~90\}th percentiles.  Thereby, \camready{together with one per-segment feature (\ie \feature{\% frames with objects})}, each video segment can be represented by 43 content features.

For each primitive, we then select the subset of features \htedit{(from the candiate set)} that correlate with its strategies.
Specifically, we pick features that have strong correlations (over 0.3 absolute Pearson correlation, a threshold suggested in~\cite{mukaka2012guide}) with at least one strategy of the four \vps studied in \S\ref{sec:motivation}. \htedit{Here we intentionally leave out three \vps (AWStream, Reducto, DDS) and use them as a holdout to test the generalizability of \name (\S\ref{subsec:eval:profiling}).}
To avoid selecting strongly correlated features while capturing as many distinct factors as possible, we iteratively select a new feature only when it has a low correlation with those already selected.
Table~\ref{tab:feature-selection} summarizes the selected features of each primitive.
These features can characterize the \pc profiles of existing \vps at a sufficient fine granularity.
We observe only diminishing improvements with more features. 
That said, \name can be expanded with more features as more \vps are developed.

\begin{table}[t]
\small
\begin{tabular}{cc}
\hline
  \textit{Primitives}                                                  & \textit{Selected features}                                                                                                  \\
    \hline\hline
\textbf{\begin{tabular}[c]{@{}l@{}}Temporal \\ pruning\end{tabular}} & \begin{tabular}[c]{@{}l@{}}\featuresmall{\% of frames with objects}, \featuresmall{ avg. object speed}, \\ \featuresmall{ avg. confidence score}\end{tabular}                  \\ \hline
\textbf{\begin{tabular}[c]{@{}l@{}}Spatial \\ pruning\end{tabular}}  & \begin{tabular}[c]{@{}l@{}}\featuresmall{\% of frames with objects}, \featuresmall{avg. total area of objects},\\\featuresmall{10\%ile of per-object area} \end{tabular} \\\hline
\textbf{\begin{tabular}[c]{@{}l@{}}Model \\ pruning\end{tabular}}    & \featuresmall{ 10\%ile of per-object area}, \featuresmall{avg. confidence score}   \\
  \hline
\end{tabular}
\vspace{0.3cm}
\tightcaption{\htedit{\name selects a subset of features for each of the three primitives, from the 43 candidate video features.}
  \vspace{-0.4cm}}
\label{tab:feature-selection}
\end{table}

\mypara{Video selection}
For each primitive, \name selects a subset of video segments from our coverage set (\S\ref{sec:motivation}) to cover all of its feasible\footnote{Some feature value combinations may be infeasible; \eg large per-object area but small total area of objects.} feature value combinations.
We first evenly split the range of values per feature into $n=4$ {\em feature value buckets} (we use feature value and feature value bucket interchangeably).
For each combination of feature values, we pick at most $k=4$ video segments from our coverage set. $n$ and $k$ can be increased if more videos are added.
As a result, \name selects
29 minutes of videos for temporal pruning, 19 minutes for spatial pruning, and 21 minutes for model pruning.

We should stress that the goal of video selection is {\em not} to be representative of a certain scenario (in fact it includes videos from different scenarios); instead, it finds videos to cover each important feature value combinations that heavily influence \vp performance. This process enables \pc profiling which ultimately helps produce accurate performance estimation of any particular scenario and workload \htedit{(explained in \S\ref{subsec:content-correlation}).}
On the other hand, \name meets this goal with only a small fraction of the coverage video set, because there is a highly uneven distribution of content features across video segments (\eg highway traffic videos contain mostly fast objects).

 \mypara{\htedit{Potential selection bias and mitigation}}
 The features selected by \name might be biased, since we only pick the features relevant to \htedit{the existing four} \vps.  \htedit{We partially examine \name's generality by showing that it can successfully profile AWStream, the \vp held out from our feature/video selection process (\S\ref{subsec:eval:profiling}).  As future work, we plan to expand/refine \name by applying the above feature/video selection process to additional and future \vps.
}

\begin{figure*}[t]
\centering
\includegraphics[width=0.98\linewidth]{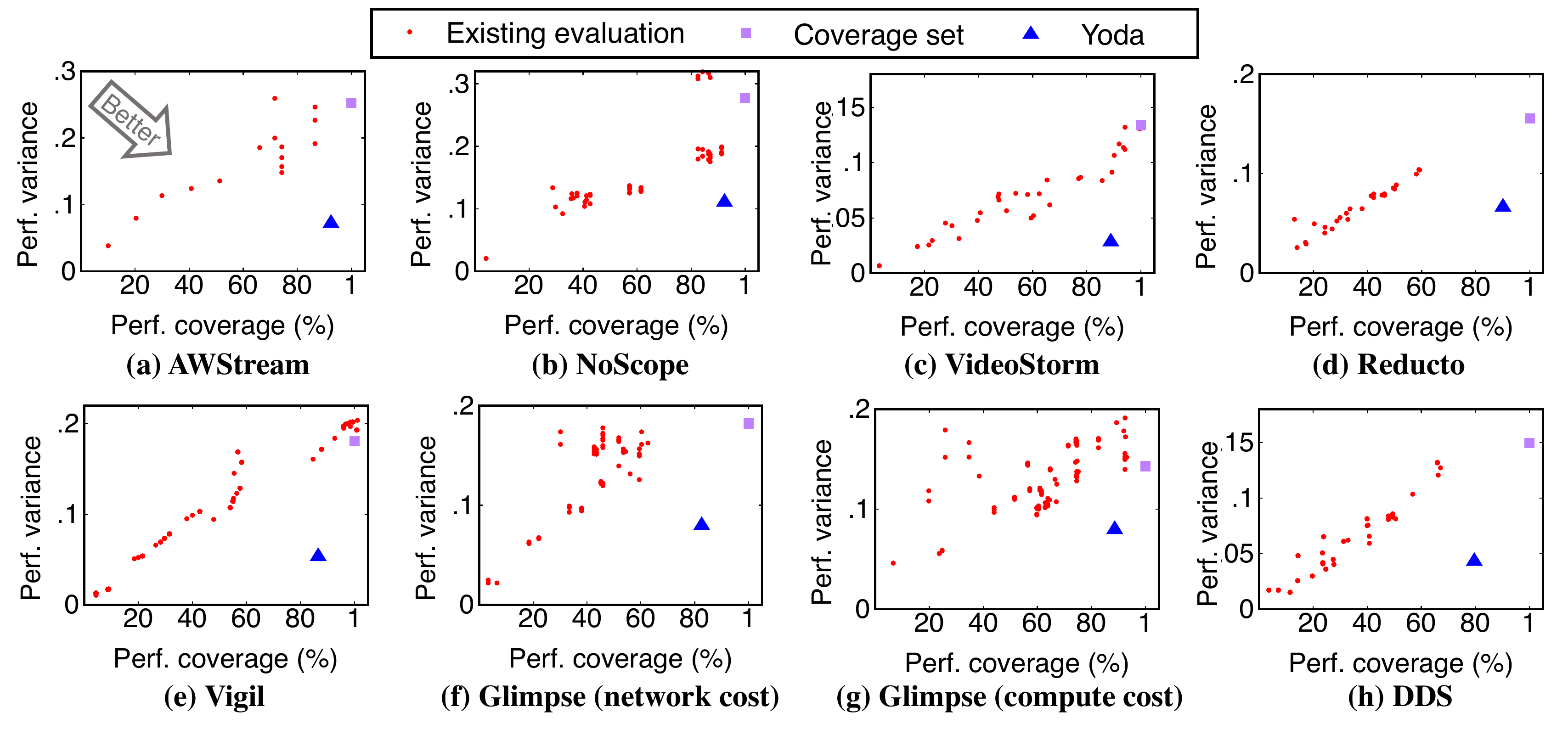}
  \vspace{-0.4cm}
\tightcaption{\name achieves a much higher level of performance
  clarity (higher coverage and lower variance), compared to existing
  evaluation methods. \camready{A high coverage means Yoda reveals both good and bad performance of a \vp, whereas a low variance means Yoda accurately estimates a \vp's performance on new videos.}
 }
\label{fig:coverage-variance}
\end{figure*}

\tightsubsection{YODA's Workflow}
\label{subsec:components}

\htedit{Using the proposed primitive-based profiling,  \name offers two key functions for its users: {\em \vp profiler} that produces a \pc profile $\Perf_\Pipeline$ for each \vp $v$, and {\em \vp performance estimator} that directly estimates $v$'s performance on a target video using $\Perf_\Pipeline$ without the need to run $v$ on the video.}

In the following, we use $\Pipeline=(\Temporal,\Spatial,\Class)$ to denote a \vp, with $\Temporal,\Spatial$ and $\Class$ being its temporal-pruning strategy, spatial-pruning strategy, and model-pruning strategy, respectively.
The \pc profile of $\Pipeline=(\Temporal,\Spatial,\Class)$ is a lookup table $\Perf_\Pipeline$ (or $\Perf_{\Temporal,\Spatial,\Class}$) that maps a feature value combination $\FeatureValVec$ in the feature space of $\FeatureVec$ to the expected performance in accuracy and cost $\Perf_{\Temporal,\Spatial,\Class}(\FeatureValVec)$.

\mypara{\vp profiler} Leveraging the property of cross-primitive independence (\S\ref{subsec:primitives}), \name builds the \pc profile of $\Pipeline=(\Temporal,\Spatial,\Class)$ in two steps.
First, we build a per-primitive profile of each of its strategies.
The temporal-pruning profile of $\Pipeline$, for instance,  is $\Perf_{\Temporal,\SpatialOracle,\ClassOracle}$,
where $\TemporalOracle,\SpatialOracle$ and $\ClassOracle$ denote the oracle strategies (see \S\ref{subsec:primitives}) of temporal pruning, spatial pruning and model pruning, respectively. \htedit{That is, we build} $\Perf_{\Temporal,\SpatialOracle,\ClassOracle}$ by setting $\Pipeline$'s spatial and model pruning strategies to their oracle ones and testing it on the benchmark videos for temporal pruning (introduced in \S\ref{subsec:features}). 
Second, we build the full \pc profile as  
\vspace{-0.1cm}
\begin{equation}
  \Perf_{\Temporal,\Spatial,\Class}(\FeatureValVec)=
\Perf_{\Temporal,\SpatialOracle,\ClassOracle}(\FeatureValVec)\cdot
\Perf_{\TemporalOracle,\Spatial,\ClassOracle}(\FeatureValVec)\cdot
\Perf_{\TemporalOracle,\SpatialOracle,\Class}(\FeatureValVec)
\end{equation}
\vspace{-0.6cm}

\mypara{\vp performance estimator}
In practice, operators often need to estimate a \vp's performance on a new (long) video.
\camready{
The challenge is that naive featurization will require annotating every object (by human annotation or running a full DNN), which can be painstakingly slow.
Fortunately, obtaining the {\em distribution} of feature values over an entire video does not require accurate results on each single frame. 
Instead, we show that running a low-cost object detector \camready{(\eg MobileNet-SSD)} on aggressively sampled frames can still yield reliable estimate of the overall feature value distribution. For instance, to get the distribution of \feature{per-object area}, we run Mobilenet-SSD on 10x uniformly sampled frames to get the area of each detected object and use the distribution of these areas as the result.
This way, \name can quickly scan a long video and produce reliable estimation of the distribution of each feature value.
}
Once the feature distribution is known, \name then uses $\Perf_\Pipeline$ to directly map the feature value distribution to $v$'s performance on the video.

\editsec{We implement \name as a ready-to-use toolkit for profiling and evaluating \vps, and plan to release the toolkit to the research community.  The toolkit provides a shared library (API) for emulating and benchmarking \vps.}


\tightsection{Evaluation}
\label{sec:usecase}
We evaluate the efficacy of \name in achieving \vp performance
clarity.  Specifically, we conduct experiments to answer the following
questions:
\begin{packed_itemize} 
\item {Does \name achieve higher \vp performance clarity, 
  compared to existing solutions that emprically test \vps on a corpus
  of videos?} (\S\ref{subsec:eval:benchmark})

\item {Is \name's primitive-based profiling accurate and efficient? Can it generalize to new \vps? } (\S\ref{subsec:eval:profiling})

 \item {Can \name accurately predict a \vp's
     performance on new videos at a low computation cost?}  (\S\ref{subsec:eval:faster-profiling})

 \item {Does \name provide new insights for \vp design and deployment?}
   (\S\ref{subsec:eval:insight}) \vspace{-0.1cm}
\end{packed_itemize}

\tightsubsection{Yoda's Performance Clarity}
\label{subsec:eval:benchmark}
%
As defined in \S\ref{subsec:content-correlation}, performance clarity
aims at providing a comprehensive characterization of \vp
performance. 
Here, we measure the level of
achieved performance clarity by two dimensions: {\em coverage} (the
completeness of the evaluation, the higher the better) and {\em
  variance} (the ambiguity of the evaluation outcome, the lower the
better). 
\camready{
The intuition is that an ideal \vp performance evaluation should have high performance coverage and low variance.
A high coverage means Yoda reveals both good and bad performance of a \vp, whereas a low variance means Yoda accurately estimates a \vp's performance on new videos.
The specific metrics of {\em coverage} and {\em variance} are defined as follows.}  Given a \vp $\Pipeline$'s PC profile
$\Perf_{\Pipeline}$ (measured from our benchmark dataset of 67-minute videos), 
\camready{\name first uses $\Perf_{\Pipeline}$ to estimate $\Pipeline$'s cost at a specific accuracy range ([0.9,0.95]) for all videos in the coverage dataset excluding our benchmark videos.}
Then, we compute 
\name's coverage as the observed cost value range, normalized by the observed cost value
range when testing $\Pipeline$ on the whole coverage dataset (14.5 hours of videos).
Next, we compute the standard deviation of \name's cost values as
\name's variance.  
Figure~\ref{fig:coverage-variance} shows the results of 
\editsec{7 pipelines in the blue boxes: \name achieves high
coverage ($>$90\%) and low variance ($<$0.2). 
The figures show one scenario per \vp, but the conclusion holds in other scenarios.
}

\mypara{\name vs. existing methods} 
Figure~\ref{fig:coverage-variance} also compares the (coverage, variance) results from the traditional evaluation method, 
which tests the performance on a long video (or a set of videos) from the target scenario (represented by the red dots).
For fairness, each test video is no shorter 
than our benchmark video.  We see that the coverage fluctuates
significantly across videos and the variance per video is much higher.
This confirms that 
\editsec{traditional} 
evaluations lead to either incomplete/partial conclusions or ambiguous
results (as we have shown in \S\ref{subsec:variability}). As a reference point, when the
\editsec{traditional} evaluation uses the entire
coverage dataset (14.5 hours), the variance exceeds 0.25, again significantly
larger than \name.

\begin{figure}[t]
\hspace*{-0.5cm}                                                           
\includegraphics[width=0.46\textwidth]{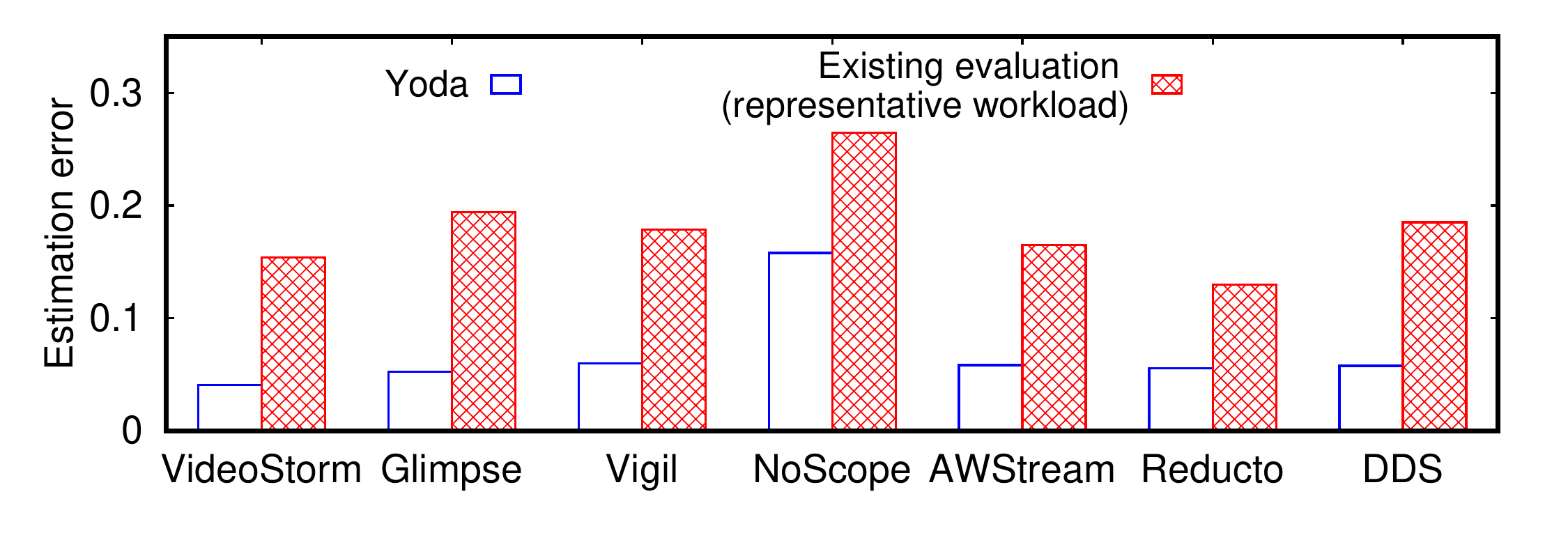}
\vspace{-0.25cm}
\tightcaption{\name provides more accurate estimation of \vp
  performance on new videos than \editsec{traditional profiling using representative workload per scenario}.
  \vspace{-0.3cm}}
\label{fig:benchmark-errors}
\end{figure}

\mypara{Microscopic study on \vp performance estimation}  We take a further step to examine the
benefit of elevated performance clarity, using the task of per-video \vp performance
estimation.  Given \name's $\Perf_{\Pipeline}$, we directly
estimate a \vp $v$'s performance on any video, and compare it to the
ground truth result obtained by running $v$ on the video.   Again we
keep the accuracy to [0.9,0.95] and measure the absolute difference
between the cost value predicted by $\Perf_{\Pipeline}$ and the ground
truth, which we refer to as cost ``estimation error''.   As reference, we apply
an ``\editsec{traditional} profiler'' to estimate $v$'s cost in the same
accuracy range by running $v$ on a representative long workload under the
same scenario of the test video, and compare it against the ground
truth.

Figure~\ref{fig:benchmark-errors} plots the median estimation errors of
both \name's profiler and the \editsec{traditional} profiler, across all the long videos in the
coverage dataset (that are not used for profiling).  We see that
\name's profiler is much more accurate than the \editsec{traditional} profiler at
estimating \vp performance on new videos.

\tightsubsection{Primitive-based Profiling: Accuracy, Cost, and Generality} 
\label{subsec:eval:profiling}

\name's efficiency partly stems from its {\em primitive-based} profiling, which tests a \vp on only videos that vary along the primitive-related features. 
To evaluate it, we compare \name with an expensive profiler built on the whole coverage set.

\mypara{Accuracy} We measure the discrepancy between the PC profile built on the whole coverage set and \name's PC profile.  The
average differences between the profiled performance curves (cost differences at same accuracy levels) are listed
in Table~\ref{tab:profile-diffs} for each of the seven \vps, and are
all very low. This corroborates our intuition in
\S\ref{subsec:features} that a small subset of videos is sufficient to profile \pc, since the feature distribution in the coverage set is highly uneven.

\mypara{Profiling cost} Profiling a \vp is a one-time cost (\ie no need to
repeat unless the \vp changes its design).   The computation cost of
profiling depends on the \vp design.  Intuitively, \vps that do not
optimize/reduce compute cost  will incur a higher overhead.  Thus we
present the result of AWStream, a \vp that does not optimize for
compute cost.  To profile AWStream, \name needs to run the \vp process
on
$\sim$72k frames using the full DNN model for object detection, at
four different video quality levels and twelve different frame
sampling rates.   When running on an Amazon EC2 machine (instance
p2.16xlarge that has 16 GPUs and costs \$14.4/hr),  the profiling
takes 8.5 minutes and cost \$2. 
\camready{Even a VAP, such as VideoStorm, that needs to profile all three primitives takes only 22.2 minutes and cost \$5.3.}

\begin{table}[t]
\footnotesize
\resizebox{0.48\textwidth}{!} {
\begin{tabular}{|>{\centering\arraybackslash}m{1.9cm}|>{\centering\arraybackslash}m{0.85cm}|>{\centering\arraybackslash}m{0.55cm}|>{\centering\arraybackslash}m{1.2cm}|>{\centering\arraybackslash}m{1.1cm}|>{\centering\arraybackslash}m{1.cm}|>{\centering\arraybackslash}m{0.85cm}|>{\centering\arraybackslash}m{0.55cm}|}
\hline
 & Glimpse & Vigil & VideoStorm & AWStream & NoScope & \editsec{Reducto} & \editsec{DDS} \\ \hline
\begin{tabular}[c]{@{}c@{}}Profile diff  (Yoda \\ vs. coverage set)\end{tabular} & 0.043 & 0.005 & 0.048 & 0.063 & 0.083 & \editsec{0.030} & \editsec{0.058} \\ \hline
\end{tabular}
}
\vspace{0.3cm}
\tightcaption{Discrepancy between the \pc profile built on \name selected videos and the \pc profile built on coverage set videos. 
}
\label{tab:profile-diffs}
\vspace{-0.65cm}
\end{table}

\mypara{Generality} 
Can \name accurately profile a new \vp not considered by
\name's feature and benchmark video selection process?  As mentioned
earlier,  we intentionally held out three of the seven \vps (AWStream, Reducto, and DDS)
from the feature/video selection process
(\S\ref{subsec:features}). Nonetheless, Figures~\ref{fig:coverage-variance} \& \ref{fig:benchmark-errors} and
Table~\ref{tab:profile-diffs} show that \name achieve similar profiling effectiveness on these holdout \vps as on other \vps.
While this does not prove that \name generalizes to all future \vps, it does indicates that \name might profile new \vps as accurately as the other \vps used in its feature selection process.

\tightsubsection{\hspace{-0.1cm}Fast-yet-accurate Performance Estimation}
\label{subsec:eval:faster-profiling}

Recall that \name offers a useful function of directly estimating a \vp $v$'s
performance on any video, without running $v$ on the video. 
We
have validated the quality of performance estimation in
Figure~\ref{fig:benchmark-errors} (\S\ref{subsec:eval:benchmark}),
using the task of estimating cost at a specific accuracy range ([0.9,0.95]). 
Below we provide more results on its estimation accuracy and computation
cost.

We consider the task to understand the variability of \vp accuracy
throughout 
a target video. For this we define two metrics on accuracy variability: 
(1) fraction of video segments whose accuracy is above 0.85, denoted by $\alpha$; and 
(2) fraction of video segments whose accuracy is below 0.7, denoted by $\beta$. 
Such metrics are useful in practice since operators often need to maintain accuracy at an acceptable level.
We use the accuracy distribution of actually running the \vp on the
video as the ground truth and define the estimation error by
$|\alpha_{estimated}-\alpha_{real}|$ and
$|\beta_{estimated}-\beta_{real}|$.

We also evaluate \name against a ``resource friendly'' baseline that actually runs $v$ on a sample set of video frames, whose 
estimation accuracy and overhead depend on the sampling rate.
Note that as explained in \S\ref{subsec:components}, \name's
performance estimator also needs to scan a sample set of video frames to
measure the video's feature value distribution. Thus its accuracy and
overhead also vary with the sampling rate.

\begin{figure}[t]
  \vspace{-0.1cm}
\centering
\begin{subfigure}[b]{0.24\textwidth}
	\centering
    \includegraphics[width=\linewidth]{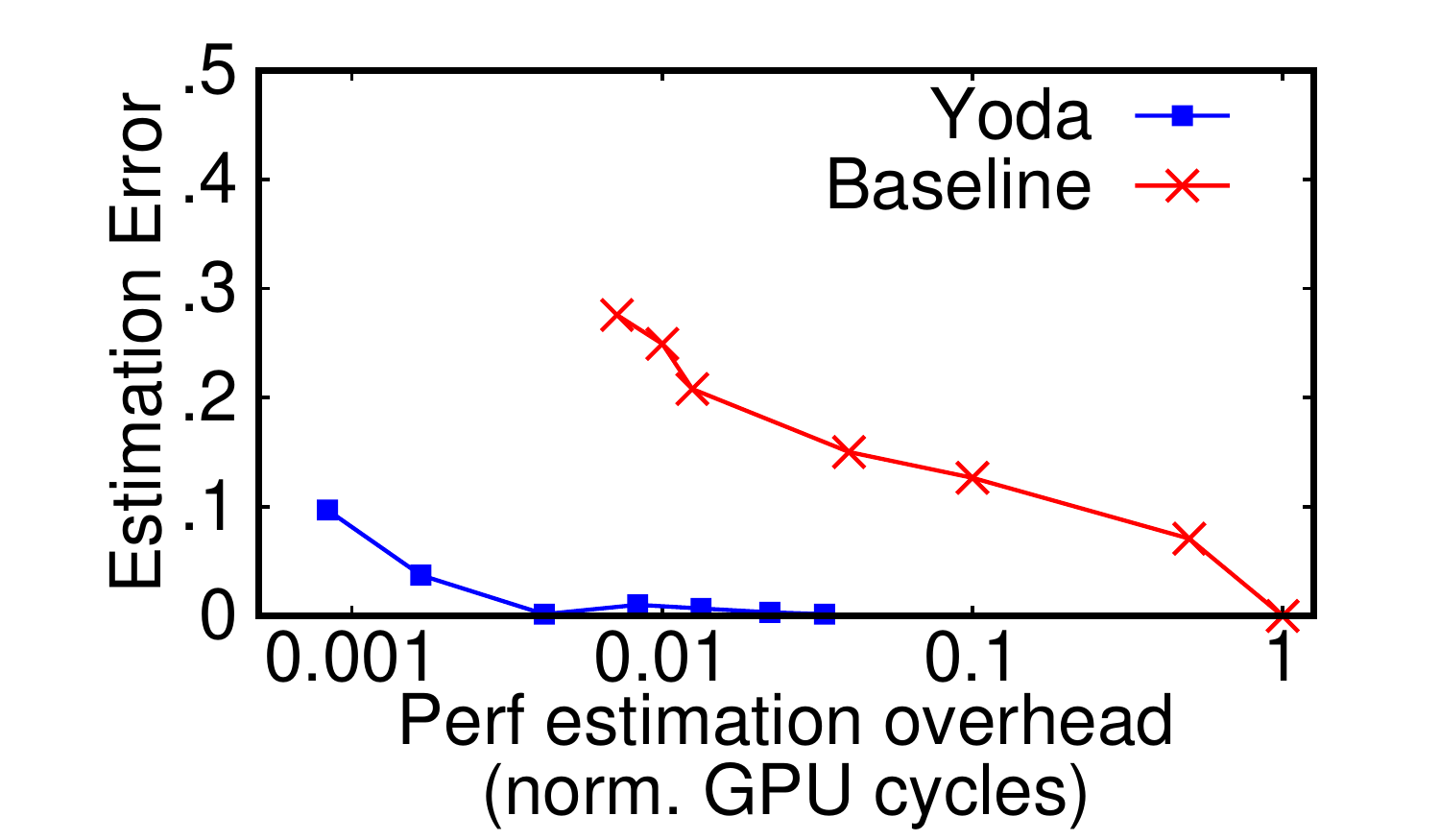}
    \subcaption{Estimating VideoStorm's fraction of high accuracy ($\alpha$)}
    \label{fig:feature-scan-videostorm-good}
\end{subfigure}
\hspace{-0.5cm}
\begin{subfigure}[b]{0.24\textwidth}
	\centering
    \includegraphics[width=\linewidth]{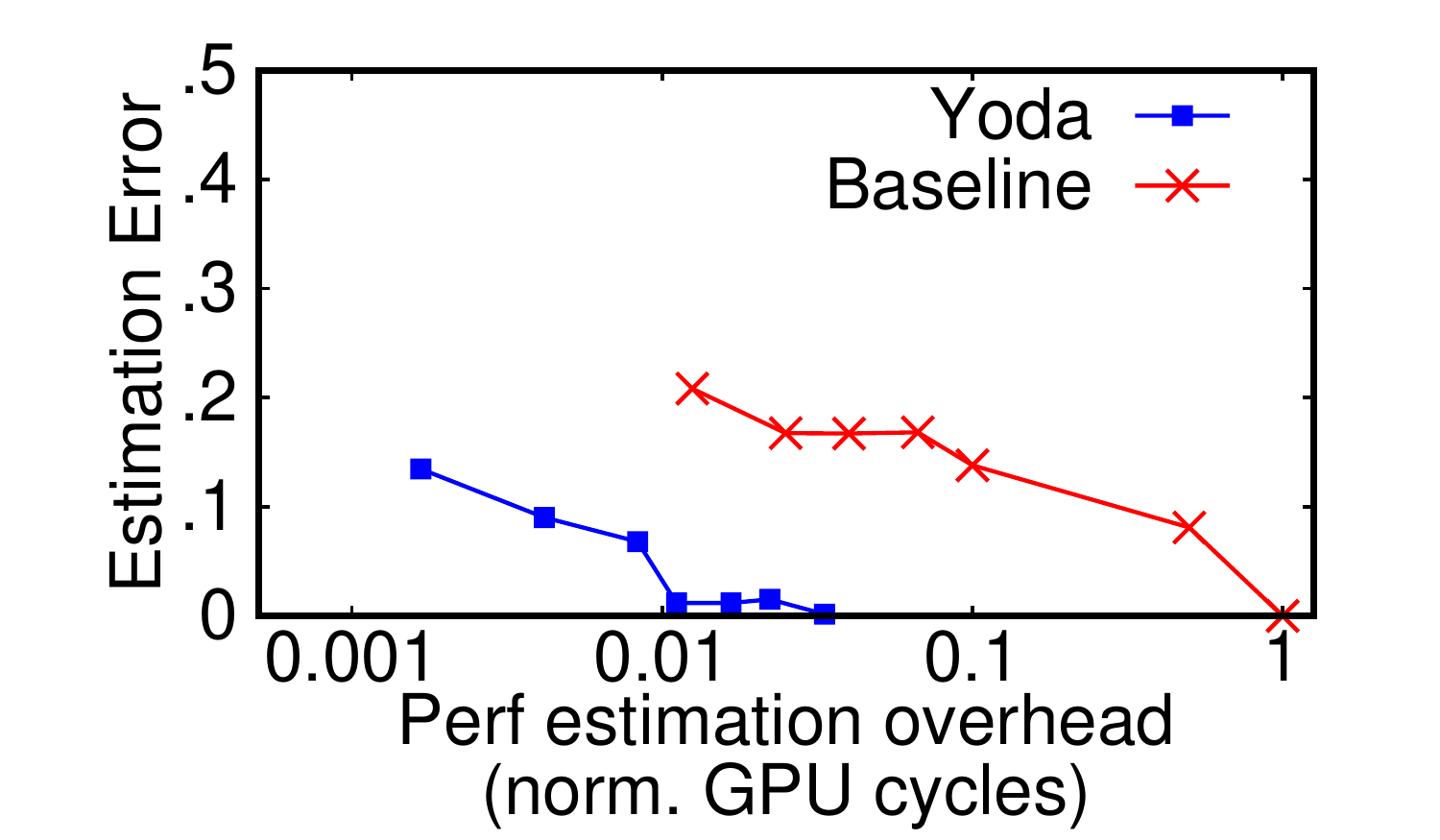}
    \subcaption{Estimating VideoStorm's fraction of low accuracy ($\beta$)}
    \label{fig:feature-scan-videostorm-bad}
\end{subfigure}
\medskip
\begin{subfigure}[b]{0.24\textwidth}
	\centering
    \includegraphics[width=\linewidth]{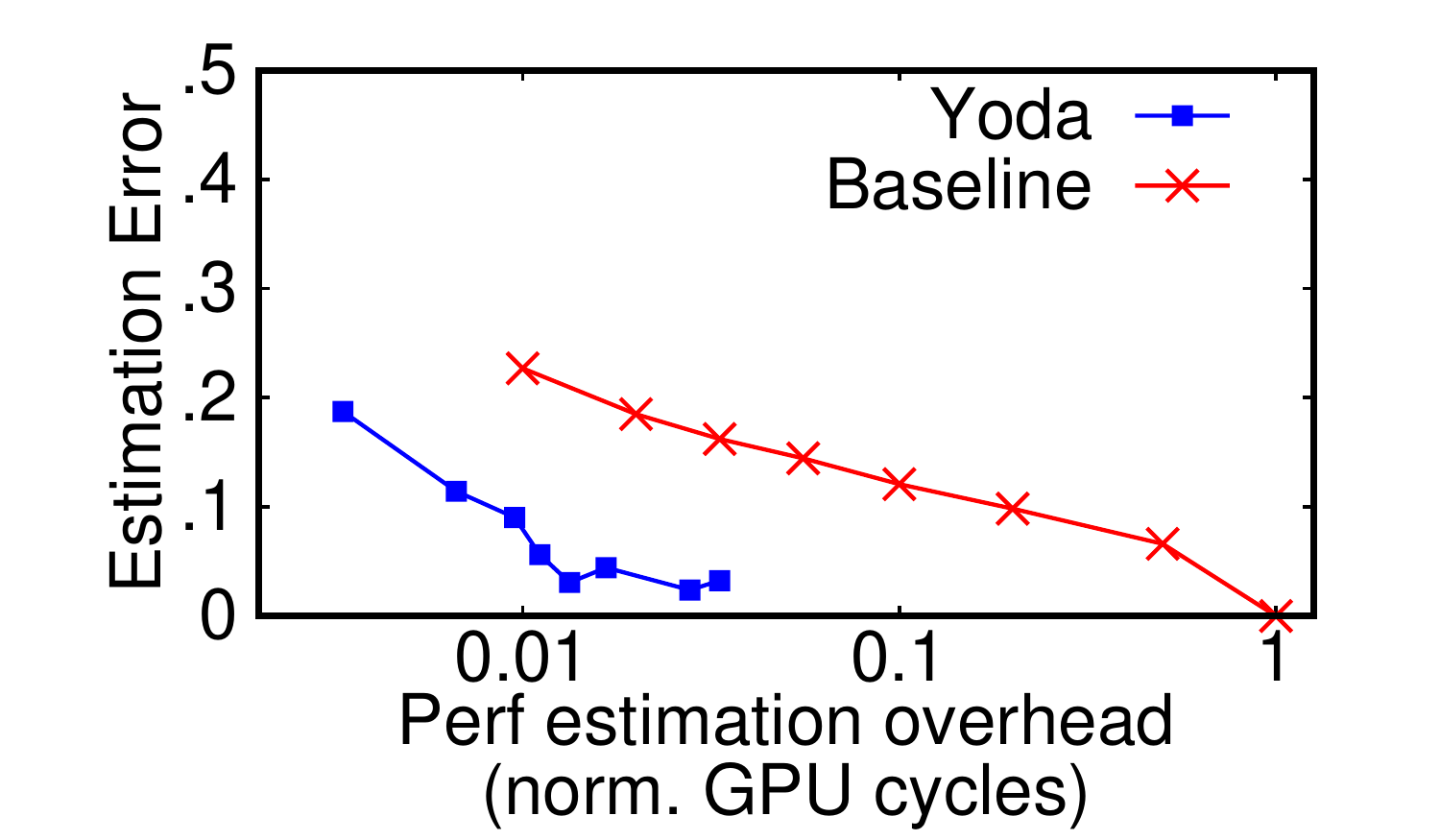}
    \subcaption{Estimating AWStream's fraction of high accuracy ($\alpha$)}
    \label{fig:feature-scan-awstream-good}
\end{subfigure}
\hspace{-0.5cm}
\begin{subfigure}[b]{0.24\textwidth}
	\centering
        \includegraphics[width=\linewidth]{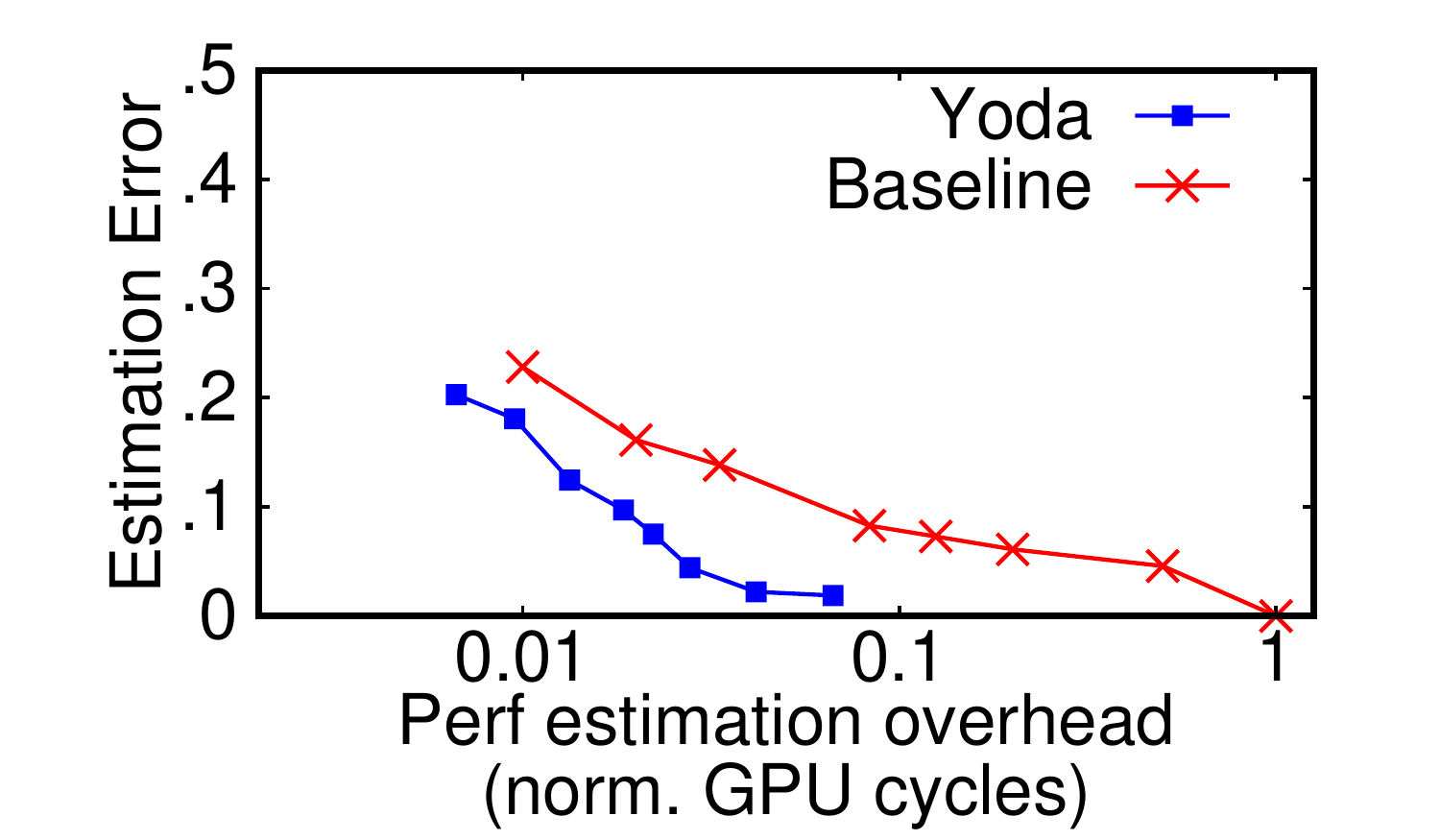}
    \subcaption{Estimating AWStream's fraction of low accuracy ($\beta$)}
    \label{fig:feature-scan-awstream-bad}
  \end{subfigure}
  \vspace{-0.2cm}
\tightcaption{\name estimates \vp performance faster and more accurately than actually running \vp on the test videos. }
  \label{fig:scanning}
  \vspace{-0.0cm}
\end{figure}

Figure~\ref{fig:scanning} shows the estimation errors of \name and baseline on
VideoStorm and AWStream, as a function of the estimation overhead
(amount of GPU cycles consumed),  for 5 hours of dashcam videos (not used during
profiling).  Here \name uses MobileNet-SSD~\cite{google-model-zoo} as the cheap object
detector to scan the videos.   For clarity, we normalize the estimation
overhead by the amount of GPU cycles consumed by running each \vp on
the full video.  We see that \name achieves nearly perfect estimation 
at a much lower cost, \ie nearly 2 orders of magnitude faster than running the
\vp on the video.

\tightsubsection{Practical Insights for \vp Deployment}
\label{subsec:eval:insight}

By providing a comprehensive profiling on \vp
performance,  \name also identifies new insights for guiding \vp design and
deployment.  We highlight two concrete use cases here.

\begin{figure}[t]
\begin{subfigure}{.15\textwidth}
  \centering
  \includegraphics[width=\linewidth]{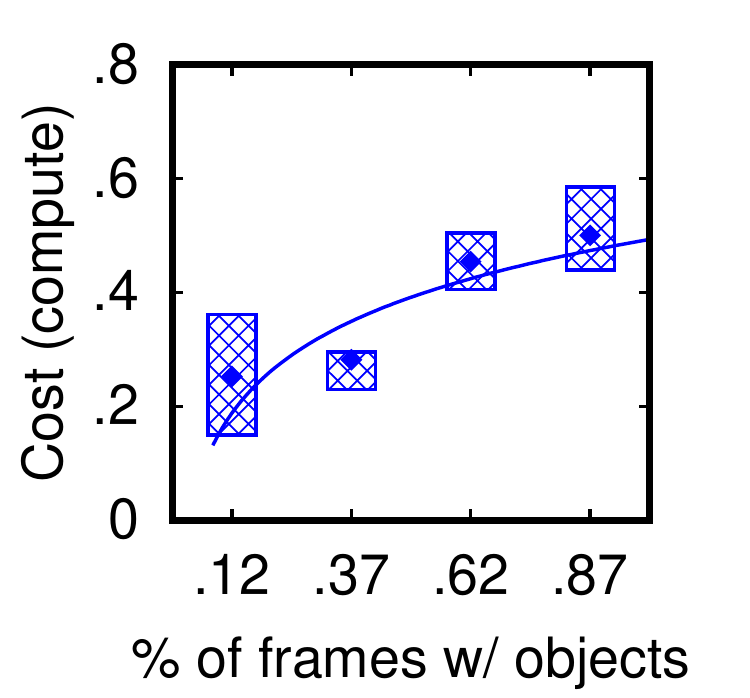}  
  \caption{Compute cost v.s. $x_1$ when $x_2$ >= 1.6.}
  \label{fig:glimpse-compute-percentage-high}
\end{subfigure}
\hfill
\begin{subfigure}{.15\textwidth}
  \centering
  \includegraphics[width=\linewidth]{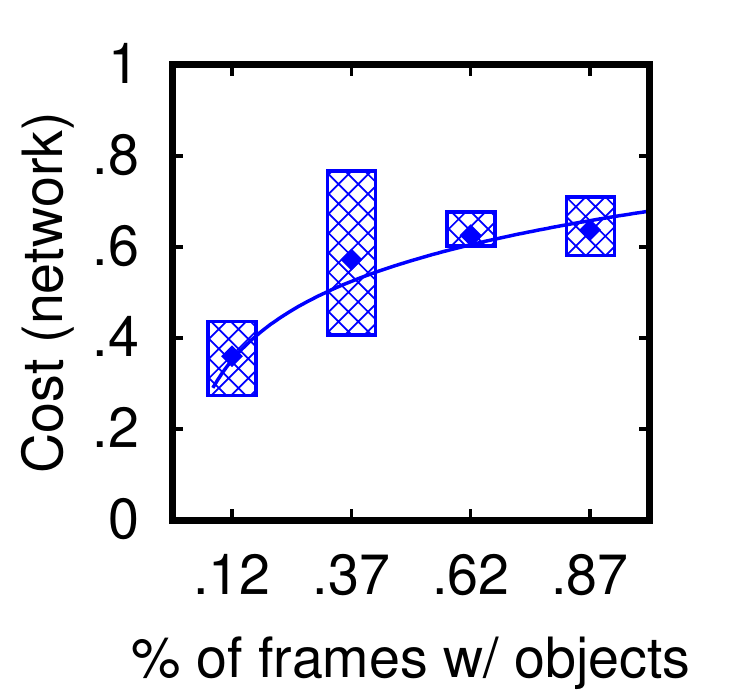}  
  \caption{Network cost v.s. $x_1$ when $x_2$ >= 1.6.}
  \label{fig:glimpse-network-percentage-high}
\end{subfigure}
\hfill
\begin{subfigure}{.15\textwidth}
  \centering
  \includegraphics[width=\linewidth]{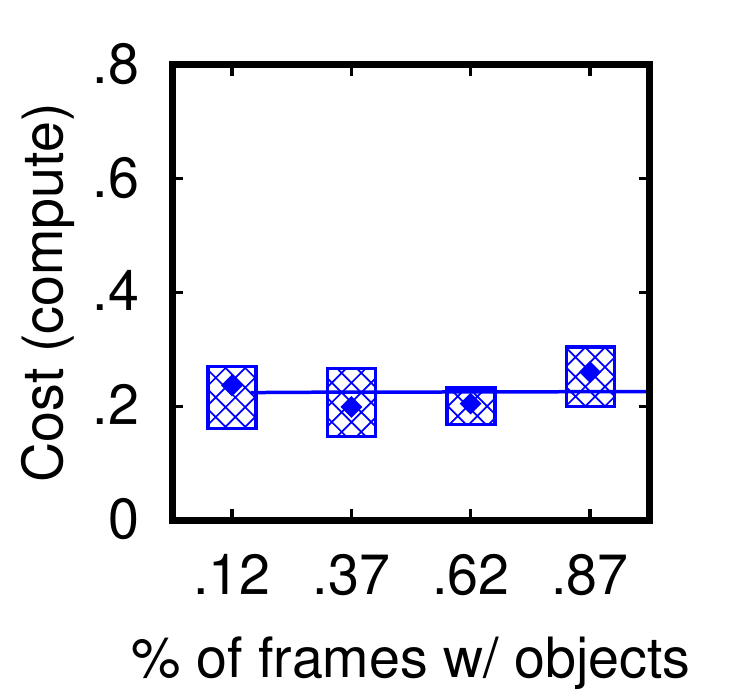}  
  \caption{Compute cost v.s. $x_1$ when $x_2$ < 1.6.}
  \label{fig:glimpse-compute-percentage-low}
\end{subfigure}
\tightcaption{Impact of feature $x_1$ (\feature{\% of frames with objects}) on performance depends on the value of feature $x_2$ (\feature{avg. object speed}).
Each box shows the mean and 25$^{\textrm{th}}$ and 75$^{\textrm{th}}$ \%iles. 
   }
   \label{fig:per-primitive-clarity}
   \vspace{-0.2cm}
\end{figure}

\mypara{Conditional correlations among features} 
Figure~\ref{fig:per-primitive-clarity} shows the performance of
Glimpse's temporal pruning strategy against two features: $x_1$ (\feature{\% of frames with objects}) and $x_2$ (\feature{average object speed}).
For better visualization, we only show the minimum cost while maintaining accuracy over
0.9 (\ie a slice of the cost-accuracy tradeoff). 
Figure~\ref{fig:per-primitive-clarity}(a) and (b) show that both
compute and network costs are strongly correlated with $x_1$ when
$x_2$ is over 1.6 (which is a
typical  vehicle speed in highway videos).
But when $x_2$ is below 1.6 (Figure~\ref{fig:per-primitive-clarity}(c)), the correlation becomes remarkably weaker.\footnote{A closer look at the selected frames shows that 
frame difference-triggered selection is no longer effective when the object speeds are so low that the frame difference triggered by their movement can easily be confused with pixel differences caused by noises in the background. }

This result implies that when testing \vps that use this pruning strategy
(\eg NoScope, Glimpse), {\em the traditional method may either
  miss this correlation} (if most test videos have slow moving objects) or {\em claim
  a strong correlation} (if most test videos have fast moving objects).
In contrast, \name reveals not only both correlation patterns, but also {\em when} they emerge, which helps to decide if a \vp should be deployed in certain video content.

\begin{figure}[t]
\hspace{-0.2cm}
\begin{subfigure}{.23\textwidth}
  \centering
  \includegraphics[width=0.9\linewidth]{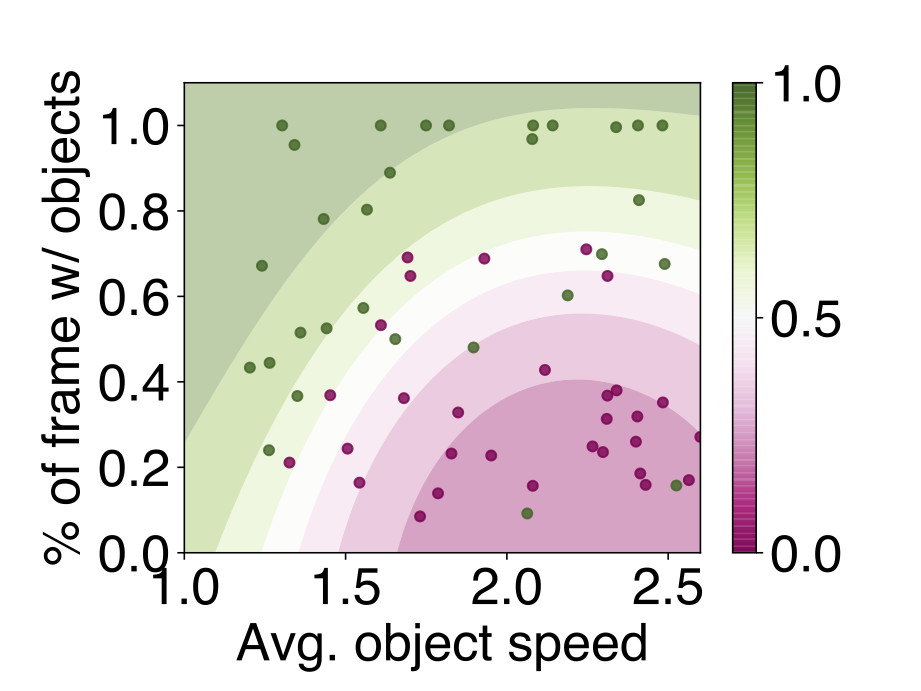}  
  \vspace{-0.2cm}
  \caption{Temporal pruning: Uniform vs. frame diff-triggered selection}
  \label{fig:awstream-temporal}
\end{subfigure}
\hspace{-0.1cm}
\begin{subfigure}{.23\textwidth}
  \centering
  \includegraphics[width=0.9\linewidth]{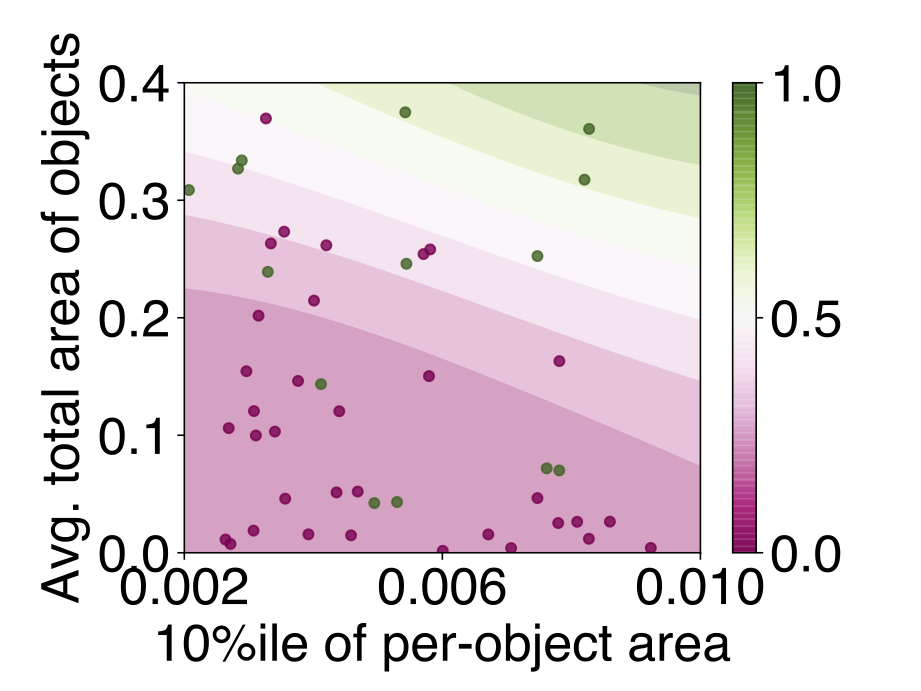}  
  \vspace{-0.2cm}
  \caption{Spatial pruning: Image quality downsizing vs. region cropping}
  \label{fig:awstream-spatial}
\end{subfigure}
\tightcaption{
In both primitives, there is no single strategy that fits in all type of content. 
The coloring indicates that where one strategy is likely better than the other.
}
\label{fig:primitive-comparison}
\end{figure}

\mypara{Informed choices of \vp strategies} 
As a case study, let us consider two temporal-pruning strategies (uniform frame selection vs. frame difference-triggered frame selection) and two spatial-pruning strategies (image quality downsizing vs. image region cropping).
Figure~\ref{fig:primitive-comparison}(a) shows the {\em operating regime} of each temporal-pruning strategy: 
frame difference-triggered selection is better when only a small fraction of frames contain objects and these objects move fast (magenta). 
Otherwise, uniform frame sampling is better (green). 
Similarly, Figure~\ref{fig:primitive-comparison}(b) shows that image quality downsizing is likely to be better if the objects are large and occupy more space in frames (green), and otherwise the image cropping strategy is better (magenta). 
These differences stem from how various strategies interact with
videos. 
For instance, image quality downsizing eliminates redundant pixels {\em in} large objects of interest (which can be detected with less pixels), whereas image cropping eliminates redundant pixels {\em outside of} objects of interest by subtracting background. 

These results have significant practical implications.
For instance, for urban traffic videos during peak hours, AWStream (uniform frame sampling and image downsizing) is better than Glimpse (frame difference-triggered frame selection), because the vehicles appear frequently and in large numbers and move slowly and often in relatively big sizes (crossroad cameras tend to be closer to the road than highway cameras), so it falls in the green regions of both graphs.
In contrast, for urban traffic videos during off-peak hours, where many large-size objects move quickly (\ie magenta in Figure~\ref{fig:primitive-comparison}(a) and green in Figure~\ref{fig:primitive-comparison}(b)), we should create a new \vp that combines Glimpse's temporal-pruning strategy and AWStream's spatial pruning strategy.


\vspace{-0.2cm}
\tightsection{Related work}
\label{sec:related}

\vspace{-4pt}
\mypara{Video analytics pipelines}
\editsec{Besides the \vps described in \S\ref{sec:background}, there are other \vps that utilize 
the same three primitives: temporal pruning (\eg~\cite{eaar,emo,clownfish,scaling,wang2018bandwidth,deepdecision}), spatial pruning (\eg~\cite{eaar,grace,catdet,deepdecision}) and model pruning (\eg~\cite{nestdnn,clownfish,approxdet,resolution,deepdecision}).}
Some work also reduces the compute/communication cost of computer-vision inference pipelines, through 
super resolution (\eg~\cite{cloudseg,shermeyer2019effects,adascale}),
splitting the DNN between camera and server (\eg~\cite{rexcam-hotmobile,ddnn,mcdnn,emmons2019cracking,hsu2019couper}),
DNN-aware cloud/edge resource scheduling (\eg~\cite{mcdnn,blazeit,hotnets,e2m,distream}), 
\editsec{cross-camera or cross-application correlations (\eg~\cite{amvp, spatula,distream,liu2019caesar,kestrel}),}
scalable data management and execution frameworks (\eg~\cite{optasia,deeplens-sanjay,scanner,edgeeye}), and 
DNN architectures tailored to balance throughput and accuracy (\eg~\cite{yolov3,huang2017multi,bolukbasi2017adaptive,wang2018skipnet,tsm,mainstream,flexdnn}). 
Many of these techniques leverage content-level characteristics, such as the ones we have discussed.
\htedit{We hope that by revealing the importance (and feasibility) of \pc, 
 future work can extend \name to support these \vps.}
\camready{While a few prior works have mentioned the issue of performance variability on some VAPs, the results were limited and only based on a handful of video features (\eg object size~\cite{dds-sigcomm}).  To the best of our knowledge, our work is the first to systematically study (using measurements \& building benchmarks) how video content features affect VAP performances.  }

\ignore{
Much effort has been made to 
reduce the computation/communication cost of computer-vision inference pipelines, through 
resolution tuning or super resolution (\eg~\cite{cloudseg,shermeyer2019effects,adascale}),
DNN-aware cloud/edge resource scheduling~\cite{mcdnn,blazeit,hotnets}, 
splitting the DNN and spanning it between the camera and the server~\cite{ddnn,mcdnn,emmons2019cracking},
scalable data management and execution frameworks (\eg~\cite{optasia,deeplens-sanjay,scanner}), and 
DNN architectures tailored to balance throughput and accuracy (\eg~\cite{yolov3,huang2017multi,bolukbasi2017adaptive,wang2018skipnet,tsm}). 
The most related research is the recent work on video analytics pipelines.
Besides the \vps
already described in \S\ref{subsec:background}, recent work has explored resolution tuning or super resolution as useful way to reduce bandwidth and storage cost~\cite{cloudseg,shermeyer2019effects,adascale}, cloud/edge resource scheduling that is aware of DNN accuracy/resource tradeoffs~\cite{mcdnn,blazeit,hotnets}, and splitting the DNN and spanning it between the camera and the server~\cite{ddnn,mcdnn,emmons2019cracking}. 
}

\vspace{-3pt}
\mypara{Edge/video analytics benchmarks}
Several benchmarks of video analytics systems have been proposed for various focuses, including throughput of video database (\eg~\cite{visual-road,scanner}), video encoding efficiency (\eg~\cite{xiph,vbench}), and shared library to implement video inference pipelines (\eg~\cite{rocket}).
\editsec{More general benchmarks catered for edge network environments are proposed as well~\cite{benchmarkingTinyML,defog,kolosov2020benchmarking}.}
Also related to \name are those benchmarking vision-task accuracies (\eg~\cite{kitti,dukemtmc}) and their tradeoffs with throughput/latency (\eg\cite{google-benchmark}).
While most benchmarks focus on average performance across images/videos, some did observe that vision models perform differently across content~\cite{objectnet} and can be sensitive to video encoding~\cite{hendrycks2019benchmarking} or training data quality~\cite{zhao2020role}. 
\name takes one  step further to systematically reveal the influence of video content features on \vp performance.
Recent efforts in computer vision similarly demonstrate that features of the test data affect the performance of a classification model (\eg~\cite{balaji2019instance,gebru2017fine}), though they focus on perturbing the features to improve model robustness whereas \name seeks to reveal the hidden relationship between \vp performance and content features.

\camready{Traditionally, the systems community has benefited from thorough performance benchmarking of data analytics systems under a wide range of workloads (\eg~\cite{armstrong2013linkbench,ycsb}), and our work is one example of this line of work in the context of video analytics. 
}






\tightsection{Conclusion}
Our work is a response to the recent trend of building efficient mobile video
analytics systems, at the expense of significant
performance variability caused by video content dependency.
We present a measurement study to shed light on this issue for
the first time, and propose the first \vp benchmark that elevates
performance clarity (how video content affects performance).  
Although \name only scratches the surface of \vp performance clarity, it is shown to be effective and
capable of identifying hidden design tradeoffs.

\begin{acks}
We thank the anonymous reviewers and our shepherd Ming Zhao for valuable feedback, Kuntai Du for providing DDS implementation and dataset, and summer intern students (Bingnan Chen, Xingcheng Yao, and Zihan Zhu) for help with \name's open-source codebase.
This work is supported in part by NSF grants(CNS-1901466,CNS-1949650 and CNS-1923778) and CERES Center. 
Junchen Jiang is also supported by a Google Faculty Research Award. 
Any opinions, findings, and conclusions or recommendations expressed in this material are those of the authors and do not necessarily reflect the views of any funding agencies.
\end{acks}

\bibliographystyle{acm}
\bibliography{benchmarking}

\end{document}